\documentclass{JHEP3} 
\usepackage{epsfig,graphicx}
\usepackage{amssymb}
\usepackage{eepic}
\usepackage{eepicsup}
\newcommand{\hmu}{\hat{\mu}}

\newcommand{\bref}[1]{(\ref{#1})}
\def\tr{{\rm ~tr}\,}

\title{Novel Approach to Super Yang-Mills Theory on Lattice\\
 {- Exact fermionic symmetry and ``Ichimatsu'' pattern -}}

\author{Katsumi\ Itoh, Mitsuhiro\ Kato,$^a$~~Hideyuki\ Sawanaka,$^b$
Hiroto\ So$^b$ and Naoya\ Ukita$^b$\\
Faculty of Education, Niigata University, Niigata 950-2181, Japan\\
$^a$ Institute of Physics, University of Tokyo, Komaba, Meguro-ku, 153-8902 Tokyo, Japan\\
$^b$ Department of Physics, Niigata University, Niigata 950-2181,
Japan\\

	E-mail: \email{itoh@ed.niigata-u.ac.jp},
	\email{kato@hep1.c.u-tokyo.ac.jp},
	\email{hide@muse.sc.niigata-u.ac.jp}, 
	\email{so@muse.sc.niigata-u.ac.jp},
	\email{ukita@muse.sc.niigata-u.ac.jp}}
\preprint{\heplat{0210049}\\	% OR: \preprint{Aaaa/Mm/Yy\\Aaa-aa/Nnnnnn}
NIIG-DP-02-6\\	  	% Use \hepth etc. also in bibliography.  
UT-Komaba/02-12}

\abstract{We present a lattice theory with an exact fermionic symmetry,
which mixes the link and the fermionic variables.  The staggered
fermionic variables may be reconstructed into a Majorana fermion in the
continuum limit.  The gauge action has a novel structure.  Though it is
the ordinary plaquette action, two different couplings are assigned in the
``Ichimatsu pattern'' or the checkered pattern.  In the naive continuum
limit, the fermionic symmetry survives as a continuum (or an $O(a^0)$)
symmetry.  The transformation of the fermion is proportional to the
field strength multiplied by the difference of the two gauge couplings
in this limit.  This work is an extension of our recently proposed cell
model toward the realization of supersymmetric Yang-Mills theory on
lattice.}


\begin{document}
%%%%%%%%%%%%%%%%%%%%%%%%%%%%%%%%%%%%%%%%%%%%%%%%%%%%%%%%%%%%%%%%%%%%%%
%%%%%%%%%%%%%%%%%%%%%%%%%%%%%%%%%%%%%%%%%%%%%%%%%%%%%%%%%%%%%%%%%%%%%%
\section{Introduction}
%%%%%%%%%%%%%%%%%%%%%%%%%%%%%%%%%%%%%%%%%%%%%%%%%%%%%%%%%%%%%%%%%%%%%%

The importance to formulate the super Yang-Mills theory (SYM) on lattice
cannot be overemphasized\cite{Feo}.  Most of previous
approaches\cite{Montvay}-\cite{kogut} to the problem are based on the
idea of Curci-Veneziano\cite{cv}.\footnote{Recently an interesting
approach to the lattice SUSY is proposed based on the hamiltonian
formalism\cite{newKaplan}.}  According to them, theories do not have any
fermionic symmetry at a finite lattice constant and the supersymmetry
shows up only in the continuum limit.  In the absence of an exact
symmetry, one must assume a supersymmetric fixed point and investigate
the Ward-Takahashi identity for SUSY\cite{Montvay}, \cite{Taniguchi}.

In our previous paper\cite{1st}, we introduced the lattice gauge model
with an exact fermionic symmetry.  The model will be called the cell
model in the present paper.  It has a Grassmann variable on each site
and plaquettes distributed in the Ichimatsu pattern or the checkered
pattern.  Assuming the form of the preSUSY transformation for the
fermionic and link variables, we derived the relations among the
transformation parameters and coefficients in the action.  We showed
that the relations may be solved by independent variables.  Therefore,
we concluded that the cell model has the exact fermionic symmetry for a
finite lattice spacing.

In order to realize SYM on lattice, there remain several important
questions to be resolved.  First of all, we would like to recover the
spinor structure in the fermionic sector.  Second, the role of the
peculiar pattern of plaquette distribution must be clearly explained.
Third, we would like to see the relation between the continuum SUSY and
the exact fermionic symmetry.  In the present paper, we would like to
address some of the questions listed above.

The peculiar pattern of the plaquettes showed up in the construction of
the cell model\cite{1st}.  However it was not clear whether this pattern
is necessary or not for having the fermionic symmetry.  Actually, in the
present paper, we will show the presence of other two models with an
exact fermionic symmetry.  Our new models differ from the cell model
only in the gauge sector.  In our first new model, we exchange the
allowed and discarded plaquettes in the cell model.  This complimentary
pattern, when realized in three dimension, looks like a set of connected
pipes (cf. Fig.2).  So the model will be called the pipe model.
Repeating the similar procedure to our previous paper\cite{1st}, we show
the presence of the preSUSY in sect.~3.  The gauge action of the other
model is the weighted sum of those of the cell and pipe models.  So we
call it the mixed model.

There has been a question of a proper continuum limit posed for the cell
model: ie, when we switch off the fermion variables, the plaquettes are
completely dissolved.  In other words, the perturbative picture is not
present in the cell model and we are forced to study it in a
nonperturbative way.  As for the pipe model, we do not encounter the
difficulty described above, even when we remove the fermion variables.
However, we can show that the pipe model without fermions does not have an
appropriate continuum limit\cite{forthcoming}.  Therefore it is natural
to consider the cell and pipe mixed model to overcome this difficulty.

In defining the mixed model, we assign different coupling constant
$\beta_c$ or ${\beta_p}$ to each set of plaquettes.  When we consider
the preSUSY transformation in the mixed system, we have twice as many
parameters associated with the plaquettes compared to the cell model.
So it is not so obvious whether we may keep the fermionic symmetry or
not.  In the present paper, we answer to this question affirmatively.

We will present a naive continuum limit of the preSUSY transformation
for the fermionic variable.  There, we observe a very important result.
The transformation is proportional to the field strength, a good sign
toward the continuum SUSY.  It is also proportional to the difference
of the two coupling constants $\beta_c$ and ${\beta_p}$.  Therefore, if
we are to find the expected SUSY algebra in the continuum limit, it
would be present only when the two couplings are different.

In relation to the spinor structure, we show that our staggered fermion
action produces the Majorana fermion in the continuum limit.  So we may
say that our mixed theory is the system of the ordinary gauge (but with
two different coupling constants) plus the staggered Majorana fermions.
However, the complete realization of spinor structure still remains as
an unresolved question.  Further discussion on this and related points
will be found in the last section.  

This paper is organized as follows.  In the next section, we introduce
the cell, pipe and mixed models and derive the conditions for the
preSUSY invariance.  The conditions obtained here are solved in sect.~3.
In sect.~4 we present a brief summary of our discussion of the Majorana
staggered fermion given in appendix C.  The naive continuum limit of the
fermion transformation is to be studied in sect.~5.  There we also
explain some properties of our models as proper lattice models.  The
last section is devoted to a summary and discussion.  Four appendices
are added.  Appendix A is more detailed discussion in support for
sect.~3.  Appendix B is on the reality properties of various quantities.
It is explained how our fermion action is related to continuum Majorana
fermions in appendix C.  In the last appendix, we show, under some
conditions, the uniqueness of the models discussed in this paper.

%%%%%%%%%%%%%%%%%%%%%%%%%%%%%%%%%%%%%%%%%%%%%%%%%%%%%%%%%%%%%%%%%%%%%%
%%%%%%%%%%%%%%%%%%%%%%%%%%%%%%%%%%%%%%%%%%%%%%%%%%%%%%%%%%%%%%%%%%%%%%
\section{The cell + pipe mixed system and its fermionic symmetry}
%%%%%%%%%%%%%%%%%%%%%%%%%%%%%%%%%%%%%%%%%%%%%%%%%%%%%%%%%%%%%%%%%%%%%%
%%%%%%%%%%%%%%%%%%%%%%%%%%%%%%%%%%%%%%%%%%%%%%%%%%%%%%%%%%%%%%%%%%%%%%
%%%%%%%%%%%%%%%%%%%%%%%%%%%%%%%%%%%%%%%%%%%%%%%%%%%%%%%%%%%%%%%%%%%%%%

%%%%%%%%%%%%%%%%%%%%%%%%%%%%%%%%%%%%%%%%%%%%%%%%%%%%%%%%%%%%%%%%%%%%%%
\subsection{The cell and pipe models}
%%%%%%%%%%%%%%%%%%%%%%%%%%%%%%%%%%%%%%%%%%%%%%%%%%%%%%%%%%%%%%%%%%%%%%

In our previous paper, we presented the multi-cell model as connected
hypercubes.\footnote{The model is simply called the cell model in the
present paper.  Its three dimensional example is shown in Fig.~2.} 
Contrary to the ordinary lattice theory, not all the possible plaquettes
are included in the action.  In two dimensional model, the allowed and
discarded plaquettes form the checkered or the Ichimatsu
pattern.\footnote{The checkered pattern is traditionally called
``Ichimatsu'' in Japan.}  Even in an arbitrary space dimension, the cell
model carries the same pattern on any two dimensional surfaces
(cf. Figs.~1 and 2).

%%%%%%%%%%%%%%%%%%%%%%%%%%%%%%%%%%%%%%%%%%%%%%%%%%%%%%%%%%%%%%%%%%%%%%
%%%%%%%%%%%%%%%%%%%%%%%%%%%%%%%%%%%%%%%%%%%%%%%%%%%%%%%%%%%%%%%%%%%%%%
%%%%%%%%%%%%%%%%%%%%%%%%%%%%%%%%%%%%%%%%%%%%%%%%%%%%%%%%%%%%%%%%%%%%%%
\begin{figure}
{\setlength{\unitlength}{0.15mm}
 \begin{picture}(230,250)
  \allinethickness{0.4mm}
\put(0,90){
  \put(90,100){\line(1,0){120}}
  \put(90,100){\line(0,0){120}}
  \put(90,100){\line(-1,-1){78}}
  \put(215,100){\vector(1,0){0}}
  \put(90,225){\vector(0,1){0}}
  \put(10,20){\vector(-1,-1){0}}
  \shade[0.3]
  \put(0,0){\path(130,140)(170,140)(170,180)(130,180)(130,140)}
  \shade[0.3]
  \put(0,0){\path(62,112)(34,84)(34,124)(62,152)(62,112)}
  \shade[0.3]
  \put(0,0){\path(102,72)(142,72)(114,44)(74,44)(102,72)}
  \allinethickness{0.6mm}
  \shade[0.3]
  \put(0,0){\path(90,100)(90,140)(130,140)(130,100)(90,100)}
  \shade[0.3]
  \put(0,0){\path(90,100)(62,72)(62,112)(90,140)(90,100)}
  \shade[0.3]
  \put(0,0){\path(90,100)(130,100)(102,72)(62,72)(90,100)}
  \thicklines
  \put(90,180){\line(1,0){100}}
  \put(90,140){\line(1,0){100}}
  \put(62,72){\line(1,0){100}}
  \put(34,44){\line(1,0){100}}
  \put(34,44){\line(0,1){100}}
  \put(62,72){\line(0,1){100}}
  \put(130,100){\line(0,1){100}}
  \put(170,100){\line(0,1){100}}
  \put(170,100){\line(-1,-1){70}}
  \put(130,100){\line(-1,-1){70}}
  \put(90,140){\line(-1,-1){70}}
  \put(90,180){\line(-1,-1){70}}
  \put(-10,10){\large $x_1$}
  \put(220,95){\makebox(10,10){\large $x_2$}}
  \put(95,230){\makebox(10,10){\large $x_3$}}
\put(10,-25){\footnotesize (a) Ichimatsu patterns}
%\put(70,-50){\footnotesize for the cell model}
}
 \end{picture}
}
%\end{eqnarray*} 
%%%%%%%%%%%%%%%%%%%%%%%%%%%%%%%%%%%%%%%%%%%%%%%%%%%%%%%%%%%%%%%%%%%%%%
\hspace{1cm}
%%%%%%%%%%%%%%%%%%%%%%%%%%%%%%%%%%%%%%%%
%%%  Figure (3-Dim. Multi Cell)      %%%
%%%%%%%%%%%%%%%%%%%%%%%%%%%%%%%%%%%%%%%%
{\setlength{\unitlength}{0.4mm}
\begin{picture}(100,140)
 \thicklines
 %
% \shade[0.3]\put(20,20){\path(0,0)(0,20)(8,32)(28,32)(28,12)(20,0)(0,0)}
 \shade[0.3]\put(20,60){\path(0,0)(0,20)(8,32)(28,32)(28,12)(20,0)(0,0)}
 \shade[0.3]\put(48,52){\path(0,0)(0,20)(8,32)(28,32)(28,12)(20,0)(0,0)}
 \shade[0.3]\put(60,20){\path(0,0)(0,20)(8,32)(28,32)(28,12)(20,0)(0,0)}
 \shade[0.3]\put(76,84){\path(0,0)(0,20)(8,32)(28,32)(28,12)(20,0)(0,0)}
 %
% \put(48,52){\line(-2,-3){14}}
 \put(48,52){\line(-2,-3){13}}
 \put(76,64){\line(1,0){25}}
 \put(56,84){\line(0,1){25}}
 \put(34,31){\vector(-2,-3){0}}
 \put(106,64){\vector(1,0){0}}
 \put(56,114){\vector(0,1){0}}
 \put(24,26){\large $x_1$}
 \put(109,61){\large $x_2$}
 \put(59,117){\large $x_3$}
 %
% \put(20,20){
% \put(0,0){\path(0,0)(20,0)(20,20)(0,20)(0,0)}
% \put(0,0){\path(20,20)(28,32)}}
 %
 \put(60,20){\put(0,0){\path(0,0)(20,0)(20,20)(0,20)(0,0)}
 \put(0,0){\path(20,20)(28,32)}}
 \put(48,52){\put(0,0){\path(0,0)(20,0)(20,20)(0,20)(0,0)}
 \put(0,0){\path(20,20)(28,32)}}
 \put(76,84){\put(0,0){\path(0,0)(20,0)(20,20)(0,20)(0,0)}
 \put(0,0){\path(20,20)(28,32)}}
 \put(20,60){\put(0,0){\path(0,0)(20,0)(20,20)(0,20)(0,0)}
 \put(0,0){\path(20,20)(28,32)}}
 \put(40,0){\footnotesize (b) Cell model}
\end{picture}
}
%%%%%%%%%%%%%%%%%%%%%%%%%%%%%%%%%%%%%%%%
%%%%%%%%%%%%%%%%%%%%%%%%%%%%%%%%%%%%%%%%%%%%%%%%%%%%%%%%%%%%%%%%%%%%%%%%%
\hspace{1cm}
%%%%%%%%%%%%%%%%%%%%%%%%%%%%%%%%%%%%%%%%%%%%%%%%%%%%%%%%%%%%%%%%%%%%%%
%%%%%%%%%%%%%%%%%%%%%%%%%%%%%%%%%%%%%%%%%%%%%%%%%%%%%%%%%%%%%%%%%%%%%%
%%%%%%%%%%%%%%%%  The Pipe   %%%%%%%%%%%%%%%%%%%%%%%%%%%%%%%%%%%%%%%%%%%%
%%%%%%%%%%%%%%%%%%%%%%%%%%%%%%%%%%%%%%%%%%%%%%%%%%%%%%%%%%%%%%%%%%%%%%%%%
%%%%%%%%%%%%%%%%%%%%%%%%%%%%%%%%%%%%%%%%%%%%%%%%%%%%%%%%%%%%%%%%%%%%%%%%%
{\setlength{\unitlength}{0.2mm}
\begin{picture}(360,280)
 \thicklines
 \put(0,0){
 \put(80,100){\line(-1,-1){20}}
 \put(200,140){\line(1,0){40}}
 \put(120,220){\line(0,1){40}}
 \put(58,78){\vector(-1,-1){0}}
 \put(245,140){\vector(1,0){0}}
 \put(120,265){\vector(0,1){0}}
 \put(38,68){\large $x_1$}
 \put(250,135){\large $x_2$}
 \put(125,270){\large $x_3$}
 \put(100,200){\line(1,0){40}}
 \put(100,200){\line(1,1){20}}
 \put(140,200){\line(1,1){20}}
 \put(120,220){\line(1,0){40}}
 \put(100,160){\line(0,1){40}}
 \put(140,200){\line(0,-1){80}}
 \put(160,180){\line(0,1){40}}
 \put(140,160){\line(-1,0){40}}
 \put(140,160){\line(1,0){40}}
 \put(160,180){\line(-1,-1){40}}
 \put(160,180){\line(-1,-1){40}}
 \put(120,100){\line(1,1){20}}
 \put(120,100){\line(0,1){40}}
 \put(120,100){\line(-1,0){40}}
 \put(80,140){\line(1,0){40}}
 \put(80,140){\line(0,-1){40}}
 \put(80,140){\line(1,1){20}}
 \put(200,180){\line(-1,0){40}}
 \put(200,180){\line(0,-1){40}}
 \put(200,180){\line(-1,-1){20}}
 \put(180,120){\line(-1,0){40}}
 \put(180,120){\line(0,1){40}}
 \put(180,120){\line(1,1){20}}
 \put(100,120){\line(-1,-1){20}}
 \put(100,120){\line(1,0){20}}
 \put(100,120){\line(0,1){20}}
 \put(200,140){\line(-1,0){20}}
 \put(120,220){\line(0,-1){20}}
 \put(80,50){\footnotesize (c) Pipe model}
 }
\end{picture}}
%%%%%%%%%%%%%%%%%%%%%%%%%%%%%%%%%%%%%%%%%%%%%%%%%%%%%%%%%%%%%%%%%%%%%%%%%
\caption{The Ichimatsu pattern and its relation to the cell and pipe models}
\end{figure}
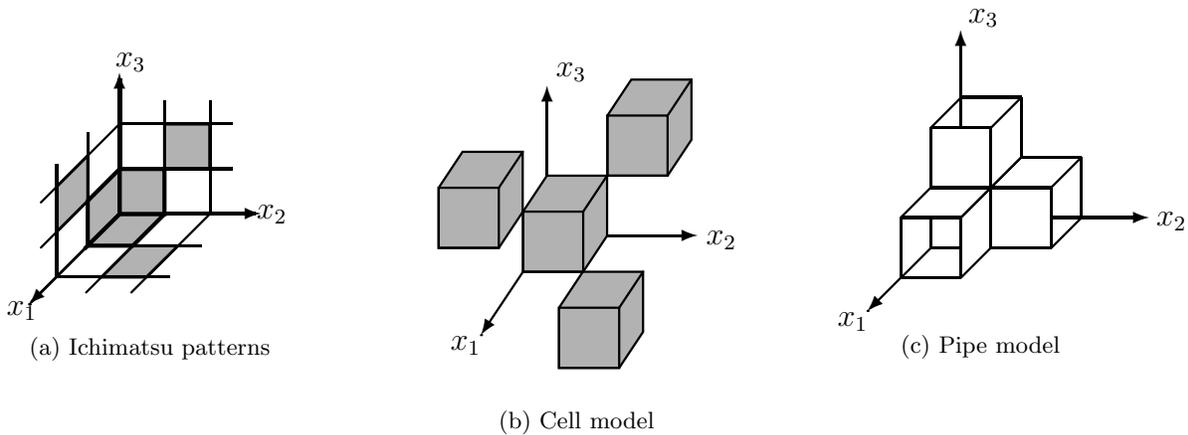
%%%%%%%%%%%%%%%%%%%%%%%%%%%%%%%%%%%%%%%%%%%%%%%%%%%%%%%%%%%%%%%%%%%%%%%%%
%%%%%%%%%%%%%%%%%%%%%%%%%%%%%%%%%%%%%%%%%%%%%%%%%%%%%%%%%%%%%%%%%%%%%%%%%
%%%%%%%%%%%%%%%%%%%%%%%%%%%%%%%%%%%%%%%%%%%%%%%%%%%%%%%%%%%%%%%%%%%%%%%%%
%%%%%%%%%%%%%%%%%%%%%%%%%%%%%%%%%%%%%%%%%%%%%%%%%%%%%%%%%%%%%%%%%%%%%%%%%
Here we show that, using the Ichimatsu pattern, we may construct another
model, called the pipe model.  It differs from the cell model only in
the gauge sector.  The gauge action includes the set of plaquettes which
are complimentary to the cell model.

In order to explain the new model, we take the three dimensional
case shown in Figs.~1 and 2 for convenience.  Consider the Ichimatsu
patterns given in Fig.~1(a).  It is easy to see that the cell model is
obtained based on the shaded pattern.  Take the shaded pattern on the
$x_1 x_2$ plane and put the same pattern on every two dimensional
surfaces parallel to it; repeat the same for the $x_2 x_3$ and $x_3 x_1$
planes; then, we obtain the cell model in Fig.~1(b).  It is also easy to
understand that, doing the same for the unshaded one, we obtain the
model shown in Fig.~1(c).  The figure looks like a connected pipe-like
object (see Fig.~1(c)).  So let us call this model the pipe model.  By
construction, allowed plaquettes for the pipe model form a complimentary
set to that of the cell model.  In Fig.~2, we show how the cell and pipe
models look like.

Starting from Fig.~1(a), we obtained the cell and pipe models.  Of
course, Fig.~1(a) is not the only way to put the Ichimatsu patterns on
the three planes; we could have started from the pattern with shaded and
unshaded plaquettes exchanged on the $x_1x_2$ plane, for example.
However, later in sect.~5, we will find that the pattern in Fig.~1(a) is
uniquely figured out once we require some properties like the rotational 
invariance.

Though explained for the three dimensional case, clearly the pipe
model can be defined for any dimension.  Its plaquettes are arranged in
the complimentary pattern to that for the cell model.

%%%%%%%%%%%%%%%%%%%%%%%%%%%%%%%%%%%%%%%%
%%%%%%%%%%%%%%%%%%%%%%%%%%%%%%%%%%%%%%%%
\begin{figure}
\hspace{1cm}
\includegraphics[scale=0.5]{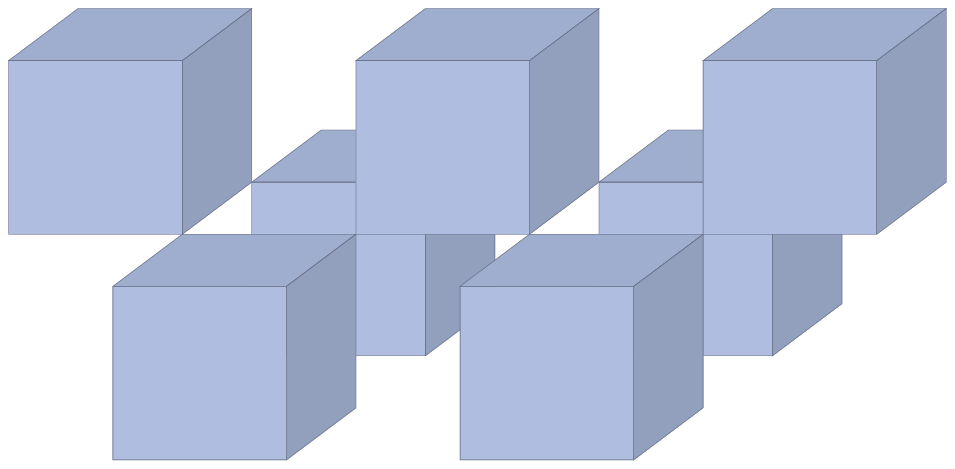}
\hspace{1cm}
\includegraphics[scale=0.5]{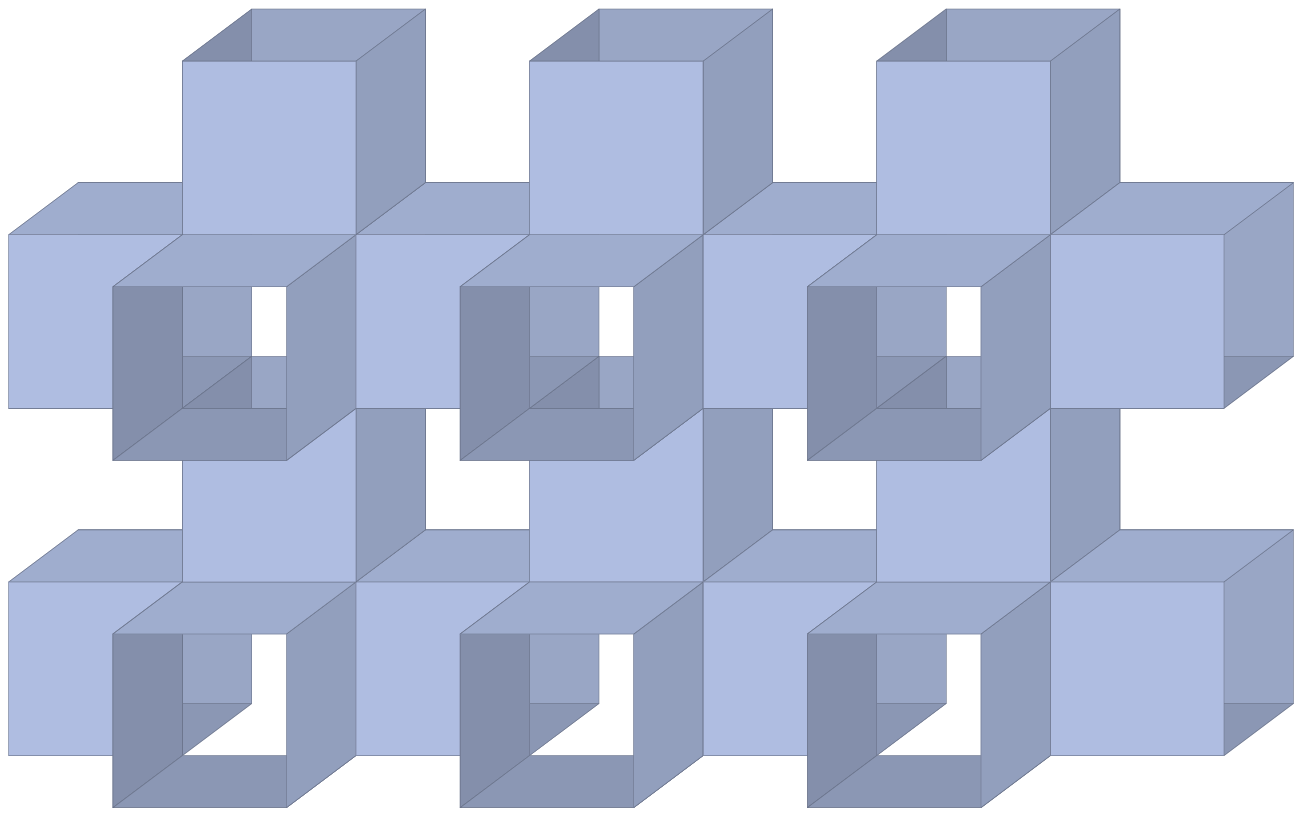}
\caption{The cell and pipe models.}
\end{figure}
%%%%%%%%%%%%%%%%%%%%%%%%%%%%%%%%%%%%%%%%
%%%%%%%%%%%%%%%%%%%%%%%%%%%%%%%%%%%%%%%%

%%%%%%%%%%%%%%%%%%%%%%%%%%%%%%%%%
%%%%%%%%%%%%%%%%%%%%%%%%%%%%%%%%%
%%%%%%%%%%%%%%%%%%%%%%%%%%%%%%%%%
\begin{floatingfigure}
%\begin{center}
%\mbox{
%%%%%%%%%%%%%%%%%%%%%%%%%%%%%%%%%%%%%%%%
%%%  Figure (Four Type Plaquettes)   %%%
%%%%%%%%%%%%%%%%%%%%%%%%%%%%%%%%%%%%%%%%
{\setlength{\unitlength}{0.2mm}
\begin{picture}(200,180)
 \thicklines
\put(20,0){ 
%%%
%%%
 \put(150,10){\line(0,1){70}}  \put(150,10){\line(-1,0){70}}
 \put(10,150){\line(0,-1){70}} \put(10,150){\line(1,0){70}}
 \shade[0.3]\put(10,10){\path(0,0)(70,0)(70,70)(0,70)(0,0)}
 \shade[0.3]\put(80,80){\path(0,0)(70,0)(70,70)(0,70)(0,0)}
 \put(95,130){\footnotesize $\epsilon = +1$}
 \put(20,25){\footnotesize $\epsilon = -1$}
 \put(15,130){\footnotesize $\epsilon = -1$}
 \put(90,25){\footnotesize $\epsilon = +1$}
 \put(150,80){\vector(1,0){40}}
 \put(80,150){\vector(0,1){40}}
 \put(65,85){\footnotesize $n$}
 \put(180,60){\footnotesize $\mu ~(< \nu)$}
 \put(90,180){\footnotesize $\nu$}
 %\put(5,190){\footnotesize $\delta = +1$}
 %\put(0,170){\footnotesize $(\delta = -1)$}
 \put(80,80){\circle*{6}}
 %
 %\put(100,90){\line(1,0){10}}
 %\qbezier(110,90)(140,90)(140,110)
 \qbezier(100,90)(142,89)(140,110)
 \put(140.7,120){\vector(0,1){0}}
 %
 %\put(60,70){\line(-1,0){10}}
 %\qbezier(50,70)(20,70)(20,50)
 \qbezier(60,70)(19,70)(20,50)
 \put(20.7,40){\vector(0,-1){0}}
 %
 %\qbezier[10](100,70)(140,70)(140,50)
 \qbezier[10](100,70)(142,71)(140,50)
 \put(140.7,40){\vector(0,-1){0}}
 %
 %\qbezier[10](50,90)(20,90)(20,110)
 \qbezier[10](60,90)(18,89)(20,110)
 \put(20.7,120){\vector(0,1){0}}
 }
\end{picture}
}
%%%%%%%%%%%%%%%%%%%%%%%%%%%%%%%%%%%%%%%%
\caption{The shaded plaquettes are for the cell model, $\delta
= +1$ and the unshaded ones for the pipe model, $\delta = -1$.  Arrows
 correspond to those in eq. \bref{plaquette} or the most rhs of eq. \bref{gauge action}. This figure is for
the case of $n_{\mu}=n_{\nu}={\rm even}$.}
\end{floatingfigure}
%%%%%%%%%%%%%%%%%%%%%%%%%%%%%%%%%
%%%%%%%%%%%%%%%%%%%%%%%%%%%%%%%%%
%%%%%%%%%%%%%%%%%%%%%%%%%%%%%%%%%

Now let us find the gauge action for the cell and pipe models.  In
Fig.~3, we see how the four plaquettes are arranged around the site $n$
on a $\mu-\nu$ plane ($\mu<\nu$).  There are always two plaquettes for
each model.  We introduce the notation $U^{\delta,
\epsilon}_{n,\mu\nu}$ to represent each open plaquette.  The index
$\delta$ is to distinguish the cell ($\delta=+1$) and pipe ($\delta=-1$)
models.  The other index $\epsilon=\pm 1$ specifies one of the two
plaquettes.  In concrete, $U^{\delta, \epsilon}_{n,\mu\nu}$ is defined
as
\begin{eqnarray}
U^{\delta, \epsilon}_{n,\mu\nu} &\equiv&
U_{n,\epsilon(-)^{n_{\mu}}\mu}
U_{n+\epsilon(-)^{n_{\mu}}{\hat \mu},\delta\epsilon (-)^{n_{\nu}}\nu}\nonumber\\
&{}& \times~U_{n+\delta\epsilon (-)^{n_{\nu}}{\hat \nu},\epsilon(-)^{n_{\mu}}\mu}^{\dagger}
U_{n,\delta\epsilon (-)^{n_{\nu}}\nu}^{\dagger}.
\label{plaquette}
\end{eqnarray}
The arrows on Fig.~3 correspond to the directions of the link variables
arranged in eq.~\bref{plaquette}, for the case of $n_{\mu}=n_{\nu}={\rm
even}$.

The gauge action may be written as
\begin{eqnarray}
S_g^{\delta} &=& -\frac{\beta_{\delta}}{2} 
\sum_{n} \sum_{0<\mu<\nu}\sum_{\epsilon=\pm 1}
{\rm tr} U^{\delta, \epsilon}_{n,\mu\nu}\nonumber\\
&=& -\frac{\beta_{\delta}}{2} \sum_{n} \sum_{0<\mu<\nu} \sum_{\epsilon=\pm 1}{\rm tr}
   %%%%%%%%%%%%%%%%%%%%
    \begin{picture}(88,28)
     \allinethickness{0.4mm}
     \put(0,0){\path(15,-12)(40,-12)(40,18)(10,18)(10,-7)}
     \thicklines
     \put(15,-12){\vector(1,0){15}}
     \put(40,-12){\vector(0,1){20}}
     \put(10,-7){\circle*{2}}
     \put(15,-12){\circle*{2}}
     \put(5,-17){\makebox(5,5){\footnotesize $n$}}
     \put(23,-23){\makebox(8,6){\footnotesize
      $\epsilon (-)^{n_{\mu}}\mu$}}
     \put(60,-2){\makebox(8,6){\footnotesize
      $\delta\epsilon (-)^{n_{\nu}}\nu$}}
    \end{picture},
   %%%%%%%%%%%%%%%%%%%%
\label{gauge action}
\end{eqnarray}
   %%%%%%%%%%%%%%%%%%%%
\\
where $\beta_{\delta=+1}=\beta_c$ and $\beta_{\delta=-1}=\beta_p$. 
Here, for later convenience, we introduced a graphical representation of
the open plaquette.  By writing $\delta$ in $S^{\delta}_g$ explicitly,
we mean that the action is for the cell {\it or} pipe model.  Later we
will introduce the cell and pipe mixed model.  But here we consider two
models separately.

We introduce non-abelian Grassmann variables $\xi \equiv \xi^a T^a$ to
represent a staggered Majorana fermion\cite{Susskind}, \cite{aratyn}:
%%%%%%%%%%%%%%%%%%%%%%%%%%%%%%%%%%%%%%%%%%%%%%%%%%%%%%%%%%%%%%%%%%%%%%
%%%%%%%%%%%%%%%%%%%%%%%%%%%%%%%%%%%%%%%%%%%%%%%%%%%%%%%%%%%%%%%%%%%%%%
 \begin{eqnarray}
  S_{f} 
 &=& \frac{1}{2} \sum_{n} \sum_{\pm\rho} b_{\pm\rho}(n) \tr
  \Bigl[ \xi_n U_{n,\pm \rho} \xi_{n \pm {\hat \rho}}
         U_{n \pm {\hat \rho},\mp \rho} \Bigr]\nonumber\\
 &=& \frac{1}{2} \sum_{n} \sum_{\pm\rho} b_{\pm\rho}(n) \tr
  \left[
   %%%%%%%%%%%%%%%%%%%%
    \begin{picture}(60,15)
     \allinethickness{0.4mm}
     \put(15,-3){\line(1,0){30}}
     \put(15,3){\line(1,0){30}}
     \put(15,3){\vector(1,0){20}}
     \put(45,-3){\vector(-1,0){20}}
     \thicklines
     \put(15,-3){\circle*{2}}
     \put(15,3){\circle*{6}}
     \put(45,0){\circle*{6}}
     \put(13,8){\makebox(5,5){\footnotesize $n$}}
     \put(30,7){\makebox(5,5){\footnotesize $\pm \rho$}}
    \end{picture}
   %%%%%%%%%%%%%%%%%%%%
  \right].
 \label{fermion action}
 \end{eqnarray}
%%%%%%%%%%%%%%%%%%%%%%%%%%%%%%%%%%%%%%%%%%%%%%%%%%%%%%%%%%%%%%%%%%%%%%
We use this fermion action for both models.  So it does not carry the
index $\delta$.  In order to have a non-vanishing action, it must hold
that $b_{-\rho}(n)= - b_{\rho}(n-{\hat \rho})$.  Later in sect.~4 we
find the expression for $b_{\rho}(n)$ which gives a Majorana fermion in
the continuum limit.  The expression satisfies this non-vanishing
condition.  In the figure on the second line of eq.~\bref{fermion
action}, the blobs represent the fermions located at the two sites,
while the lines connecting them are link variables.  Since the Grassmann
variables are present, it is important to keep the order of the
variables.  So we make the following rule: when we need to recover the equation
from a figure, we write down variables following arrows in the figure.

In sect.~4 and appendix C, we will explain how the action \bref{fermion
action} is related to the continuum Majorana fermion.

Now we would like to consider a fermionic transformation (the preSUSY
transformation) which mixes the link and the fermion variables.  We
assume the form of the transformation, hoping to realize the ordinary
SUSY in a continuum limit.  The continuum SUSY transformation of a
fermion field is proportional to the field strength.  So we take our
fermion transformation as follows,
%%%%%%%%%%%%%%%%%%%%%%%%%%%%%%%%%%%%%%%%%%%%%%%%%%%%%%%%%%%%%%%%%%%%%%
%%%%%%%%%%%%%%%%%%%%%%%%%%%%%%%%%%%%%%%%%%%%%%%%%%%%%%%%%%%%%%%%%%%%%%
 \begin{eqnarray}
  \delta \xi_n = 
  \sum_{0<\mu<\nu} \sum_{\epsilon=\pm 1} C_{n, \mu \nu}^{\delta, \epsilon}
  \left( U_{n, \mu \nu}^{\delta, \epsilon} - U_{n, \mu \nu}^{\delta, \epsilon \dagger}
  \right)
 \label{fermion tf}
 \end{eqnarray}
%%%%%%%%%%%%%%%%%%%%%%%%%%%%%%%%%%%%%%%%%%%%%%%%%%%%%%%%%%%%%%%%%%%%%%
or
%%%%%%%%%%%%%%%%%%%%%%%%%%%%%%%%%%%%%%%%%%%%%%%%%%%%%%%%%%%%%%%%%%%%%%
%%%%%%%%%%%%%%%%%%%%%%%%%%%%%%%%%%%%%%%%%%%%%%%%%%%%%%%%%%%%%%%%%%%%%%
%%%%%%%%%%%%%%%%%%%%%%%%%%%%%%%%%%%%%%%%%%%%%%%%%%%%%%%%%%%%%%%%%%%%%%
 \begin{eqnarray}
  \delta
  \left(
   %%%%%%%%%%%%%%%%%%%%
    \begin{picture}(10,9)
     \thicklines
     \put(5,3){\circle*{6}}
     \put(2,-8){\footnotesize $n$}
    \end{picture}
   %%%%%%%%%%%%%%%%%%%%
  \right) &=&
  \sum_{0<\mu<\nu} \sum_{\epsilon=\pm 1}
  \left[
   %%%%%%%%%%%%%%%%%%%%
    \begin{picture}(95,28)
     \allinethickness{0.4mm}
     \put(0,0){\path(18,-12)(45,-12)(45,18)(15,18)(15,-6)}
     \thicklines
     \put(33,-12){\vector(1,0){0}}
     \put(45,6){\vector(0,1){0}}
     \put(15,-6){\circle*{2}}
     \put(15,-12){\circle{6}}
     \put(15,-12){\makebox(0,0){$\times$}}
     %
     %%\put(5,-10){\makebox(5,5){\footnotesize $n$}}
     \put(28,-23){\makebox(8,6){\footnotesize
      $\epsilon (-)^{n_{\mu}}\mu$}}
     \put(65,-2){\makebox(8,6){\footnotesize
      $\delta\epsilon (-)^{n_{\nu}}\nu$}}
    \end{picture}
   %%%%%%%%%%%%%%%%%%%%
  -
   %%%%%%%%%%%%%%%%%%%%
    \begin{picture}(95,28)
     \allinethickness{0.4mm}
     \put(0,0){\path(56,-12)(80,-12)(80,18)(50,18)(50,-9)}
     \thicklines
     \put(70,18){\vector(1,0){0}}
     \put(50,8){\vector(0,1){0}}
     \put(50,-12){\circle{6}}
     \put(56,-12){\circle*{2}}
     \put(50,-12){\makebox(0,0){\bf $\times$}}
     % 
     %%\put(5,-10){\makebox(5,5){\footnotesize $n$}}
     \put(61,23){\makebox(8,6){\footnotesize
      $\epsilon (-)^{n_{\mu}}\mu$}}
     \put(22,-2){\makebox(8,6){\footnotesize
      $\delta\epsilon (-)^{n_{\nu}}\nu$}}
    \end{picture}
   %%%%%%%%%%%%%%%%%%%%
  \right].
 \label{fermion tf2}
 \end{eqnarray} 
%%%%%%%%%%%%%%%%%%%%%%%%%%%%%%%%%%%%%%%%%%%%%%%%%%%%%%%%%%%%%%%%%%%%%%
%%%%%%%%%%%%%%%%%%%%%%%%%%%%%%%%%%%%%%%%%%%%%%%%%%%%%%%%%%%%%%%%%%%%%%
The Grassmann transformation parameters are denoted as $C_{n, \mu
\nu}^{\delta, \epsilon}$.  In the figure they are represented by the
crossed circles.  In order to distinguish between the cell and pipe
models, this $C$-parameter carries the index $\delta$.  Later we also use
the notation $C_{n, \mu \nu}^{(\epsilon)}$ for $\delta=+1$ and
$\bar{C}_{n, \mu \nu}^{(\epsilon)}$ for $\delta=-1$.

The action is to be invariant under the preSUSY transformations of both
of the link and fermion variables.  So, once eq.~\bref{fermion tf} is
given, the form of the link variable transformation is naturally
determined.\footnote{This step is easily understood when we write
graphically the transformation of the action.}  By introducing Grassmann
odd transformation parameters $A$ and $B$, it may be written as
\begin{eqnarray}
\delta U_{n,\mu} =
(A \cdot \zeta)_{n,\mu}U_{n,\mu}
+U_{n,\mu}(B \cdot \zeta)_{n+{\hat \mu},-\mu}
\label{del U}
\end{eqnarray}
where 
\begin{eqnarray}
(A \cdot \zeta)_{n,\mu}\equiv \sum_{\pm \rho} 
A_{n,\mu}^{\pm \rho}\zeta_n^{\pm \rho},~~~~~
(B \cdot \zeta)_{n+{\hat \mu},-\mu}\equiv \sum_{\pm \rho} 
B_{n+{\hat \mu},-\mu}^{\pm \rho}\zeta_{n+{\hat \mu}}^{\pm \rho},
\label{A dot zeta}
\end{eqnarray}
and
\begin{eqnarray}
\zeta_n^{\pm \rho} \equiv U_{n,\pm \rho} \xi_{n \pm {\hat \rho}} U_{n,\pm \rho}^{\dagger}.
\label{zeta}
\end{eqnarray}

Eq.~\bref{del U} implies 
\begin{eqnarray}
\delta U_{n,\mu}^{\dagger} = - U_{n,\mu}^{\dagger} \delta U_{n,\mu} U_{n,\mu}^{\dagger} 
= -(B \cdot \xi)_{n+{\hat \mu},-\mu} U_{n,\mu}^{\dagger}
-U_{n,\mu}^{\dagger} (A \cdot \xi)_{n,\mu}.
\label{del U dagger}
\end{eqnarray}
By introducing the notation $U_{n+{\hat \mu}, -\mu} \equiv
U_{n,\mu}^{\dagger}$, we may write eqs. \bref{del U} and \bref{del U
dagger} in a single equation,
\begin{eqnarray}
\delta U_{n,\sigma \mu} =
(\alpha \cdot \zeta)_{n,\sigma \mu}U_{n,\sigma \mu}
- U_{n,\sigma \mu} ({\alpha} \cdot \zeta)_{n+\sigma {\hat \mu},-\sigma \mu}
\label{One cell del U}
\end{eqnarray}
where $\alpha$ are related to $A$ or $B$ in eqs. \bref{del U} and \bref{del U
dagger} as
\begin{eqnarray}
\alpha_{n,\mu}^{\pm \rho} = A_{n,\mu}^{\pm \rho},~~~~~\alpha_{n,-\mu}^{\pm \rho} = - B_{n,-\mu}^{\pm \rho}.\nonumber
\end{eqnarray}
We have three comments. 1) In writing \bref{One cell del U} or
originally eq.~\bref{A dot zeta}, we made a minor change of notation
from our previous paper\cite{1st}: the minus sign of $-\mu$ of
$\alpha_{n+{\hat \mu},-\mu}^{\pm \rho}$ indicates the $\alpha$-parameter
is located at the end point of the transformed link variable
$U_{n,\mu}$.  2) The relation $\delta U^{\dagger}=-U^{\dagger}\delta U
U^{\dagger}$ is obtained from the unitarity of the link variable.  The
result must be the same as $(\delta U)^{\dagger}$.  Therefore the
quantity $(\alpha \cdot \zeta)$ must be pure imaginary.  Then, each term
in \bref{One cell del U} may be regarded as an infinitesimal form of an
unitary transformation acted from the left or the right.  This fact is
important for the invariance of the path integral measure\cite{1st}. 3)
Though eq.~\bref{One cell del U} may look like a gauge transformation,
it is not so.  In the next section we solve conditions on the parameters
for the action invariance.  Then we clearly understand that the
solution, expressed in terms of independent parameters, does not contain
the gauge transformation.

For later convenience, we introduce a graphical representation of
eq.~\bref{One cell del U} as
%%%%%%%%%%%%%%%%%%%%%%%%%%%%%%%%%%%%%%%%%%%%%%
%%%%%%%%%%%%%%%%%%%%%%%%%%%%%%%%%%%%%%%%%%%%%%
\begin{eqnarray}
\delta \Bigl( 
%%%%%%%%%%%%%%%%%%%%%%%%%%%%%%%%%%%%%%%%%%%%%%
	\begin{picture}(60,30)
	\thicklines
	\put(10,3){
	\put(0,0){\vector(1,0){20}}
	\put(0,0){\line(1,0){30}}
	\put(0,0){\circle*{2}}
	\put(30,0){\circle*{2}}
	\put(-5,-12){\footnotesize $n$}
	\put(15,10){\footnotesize $\sigma \mu$}
	\put(20,-12){\footnotesize $n+\sigma {\hat \mu}$}
	}
	\end{picture}
%%%%%%%%%%%%%%%%%%%%%%%%%%%%%%%%%%%%%%%%%%%%%%
\Bigr)=
\sum_{\rho, \sigma^{\prime}}
\Big[
%%%%%%%%%%%%%%%%%%%%%%%%%%%%%%%%%%%%%%%%%%%%%%
\begin{picture}(60,35)
\thicklines
\put(5,3){
	\put(0,10){\footnotesize $\sigma^{\prime} \rho$}
	\put(16,-15){\footnotesize $n$}
	\put(30,-15){\footnotesize $\sigma \mu$}
	\put(18,-3){\circle{6}}
	\put(21,22){\circle*{6}}
	\put(18,0){\vector(0,1){12}}
	\put(18,0){\line(0,1){22}}
	\put(24,22){\vector(0,-1){18}}
	\put(24,22){\line(0,-1){25}}
	\put(24,-3){\vector(1,0){18}}
	\put(24,-3){\line(1,0){26}}
	\put(50,-3){\circle*{2}}
	}
\end{picture}
%%%%%%%%%%%%%%%%%%%%%%%%%%%%%%%%%%%%%%%%%%%%%%
-
%%%%%%%%%%%%%%%%%%%%%%%%%%%%%%%%%%%%%%%%%%%%%%
\begin{picture}(60,35)
\thicklines
\put(5,3){
	\put(12,10){\footnotesize $\sigma^{\prime} \rho$}
	\put(2,-13){\footnotesize $n$}
	\put(14,-15){\footnotesize $\sigma \mu$}
	\put(30,-3){\circle{6}}
	\put(33,22){\circle*{6}}
	\put(36,-3){\circle*{2}}
	\put(30,0){\vector(0,1){12}}
	\put(30,0){\line(0,1){20}}
	\put(36,22){\vector(0,-1){18}}
	\put(36,22){\line(0,-1){25}}
	\put(4,-3){\vector(1,0){18}}
	\put(4,-3){\line(1,0){23}}
	\put(4,-3){\circle*{2}}
	}
\end{picture}
%%%%%%%%%%%%%%%%%%%%%%%%%%%%%%%%%%%%%%%%%%%%%%
\Bigr].
\label{link tf pic}
\end{eqnarray}
%%%%%%%%%%%%%%%%%%%%%%%%%%%%%%%%%%%%%%%%%%%%%%
%%%%%%%%%%%%%%%%%%%%%%%%%%%%%%%%%%%%%%%%%%%%%%
The circles are for the $\alpha$-parameters.  

As is easily realized, the variation of the action under the preSUSY
contains terms cubic and linear in the fermion variables.  For the
action to be invariant, these two set of terms should vanish separately.

The change of the fermion action \bref{fermion action} when link
variables are transformed, denoted as $\delta_U S_f$, consists of
fermion cubed terms.  A term may be represented by the following figure,
%%%%%%%%%%%%%%%%%%%%%%%%%%%%%%%%%%%%%%%%%%%%%%%%%%%%%%%%%%%%%%%%%%%%%%
%%%%%%%%%%%%%%%%%%%%%%%%%%%%%%%%%%%%%%%%%%%%%%%%%%%%%%%%%%%%%%%%%%%%%%
 \begin{eqnarray}
\tr 
  \left[~~~
   %%%%%%%%%%%%%%%%%%%%
    \begin{picture}(60,28)
     \allinethickness{0.4mm}
     \put(0,0){\path(20,17)(20,-7)(44,-7)}
     \put(14,-13){\line(0,1){30}}
     \put(14,-13){\vector(0,1){20}}
     \put(20,-13){\line(1,0){24}}
     \put(20,-7){\vector(1,0){15}}
     \thicklines
     \put(14,-13){\circle*{6}}
     \put(17,17){\circle*{6}}
     \put(44,-10){\circle*{6}}
     \put(20,-13){\circle*{2}}
\put(5,-20){\footnotesize $n$}
\put(30,0){\footnotesize $\sigma \mu$}
\put(-5,3){\footnotesize $\sigma^{\prime} \rho$}
\put(0,24){\footnotesize $n+\sigma^{\prime} {\hat \rho}$}
\put(28,-22){\footnotesize $n+\sigma {\hat \mu}$}
    \end{picture}
   %%%%%%%%%%%%%%%%%%%%
  \right].
\label{fermion cubed term}
 \end{eqnarray}
%%%%%%%%%%%%%%%%%%%%%%%%%%%%%%%%%%%%%%%%%%%%%%%%%%%%%%%%%%%%%%%%%%%%%%
%%%%%%%%%%%%%%%%%%%%%%%%%%%%%%%%%%%%%%%%%%%%%%%%%%%%%%%%%%%%%%%%%%%%%%
%%%%%%%%%%%%%%%%%%%%%%%%%%%%%%%%%%%%%%%%%%%%%%%%%%%%%%%%%%%%%%%%%%%%%%
Here $\sigma$ and $\sigma^{\prime}$ are sign factors.  This figure may
be generated in two different ways, by the transformation of the link
variable $U_{n, \sigma \mu}$ or $U_{n+\sigma^{\prime} \rho,
-\sigma^{\prime} \rho}$.  Taking care of the sign factor due to the
Grassmann nature of the $\alpha$-parameter, we obtain
%%%%%%%%%%%%%%%%%%%%%%%%
%%%%%%%%%%%%%%%%%%%%%%%%
\begin{eqnarray}
0&=&- b_{\sigma \mu}(n) \alpha_{n,\sigma \mu}^{\sigma^{\prime} \rho}
+ b_{-\sigma^{\prime} \rho}(n+\sigma^{\prime} \rho)
\alpha_{n,\sigma^{\prime} \rho}^{\sigma \mu} \nonumber\\
&=&- b_{\sigma \mu}(n) \alpha_{n,\sigma \mu}^{\sigma^{\prime} \rho}
- b_{\sigma^{\prime} \rho}(n)
\alpha_{n,\sigma^{\prime} \rho}^{\sigma \mu}. 
\label{fermion cubed condition}
\end{eqnarray}
%%%%%%%%%%%%%%%%%%%%%%%%
%%%%%%%%%%%%%%%%%%%%%%%%
Note that eq.~\bref{fermion cubed condition} should hold both for cell and
pipe models, independently of $\delta$.

%%%%%%%%%%%%%%%%%%%%%%%%%%%%%%%%%%%%%%%%%%%%%%%%%%%%%%%%%%%%%%%%%%%%%%
%%%%%%%%%%%%%%%%%%%%%%%%%%%%%%%%%%%%%%%%%%%%%%%%%%%%%%%%%%%%%%%%%%%%%%
%\noindent
%\underline{Fermion linear $\delta_U S_g + \delta_{\xi}S_f$}\\

In a similar manner, we may study the cancellation of fermion linear terms, 
$\delta_U S_g^{\delta} + \delta_{\xi}S_f=0$,
%%%%%%%%%%%%%%%%%%%%%%%%%%%%%%%%%%%%%%%%%%%%%%%%%%%%%%%%%%%%%%%%%%%%%%
%%%%%%%%%%%%%%%%%%%%%%%%%%%%%%%%%%%%%%%%%%%%%%%%%%%%%%%%%%%%%%%%%%%%%%
\begin{eqnarray}
0&=&\sum_{n,0<\mu<\nu} \sum_{\rho,\epsilon,\sigma}
\Biggl( -\frac{1}{2} b_{\sigma \rho}(n)
\hspace{0.3cm}
\begin{picture}(50,50)
 \allinethickness{0.4mm}
\put(0,-20){
 \put(0,0){\path(30,36)(57,36)(57,30)(33,30)}
 \put(0,0){\path(30,27)(30,0)(0,0)(0,30)(24,30)}
 \thicklines
 \put(57,33){\circle*{6}}
 \put(30,36){\vector(1,0){16}}
 \put(30,27){\vector(0,-1){14}}
 \put(30,36){\circle*{2}}
 \put(24,30){\circle*{2}}
     \put(20,36){\small $n$}
     \put(40,45){\footnotesize $\sigma \rho$}
     \put(52,10){\makebox(8,6){\footnotesize
      $\epsilon (-)^{n_{\mu}}\mu$}}
     \put(20,-15){\makebox(8,6){\footnotesize
      $\delta\epsilon (-)^{n_{\nu}}\nu$}}
     \put(30,30){\circle{6}}
     \put(30,30){\makebox(0,0){$\times$}}
}
\end{picture}
\nonumber\\
&{}&-\frac{1}{2} \beta_{\delta}
\biggl(
\hspace{1cm}
\begin{picture}(50,50)
 \allinethickness{0.4mm}
\put(0,0){
 \put(0,0){\path(-6,-3)(-6,-24)}
 \put(0,0){\path(0,-24)(0,0)(30,0)(30,30)(0,30)(0,6)}
 \thicklines
 \put(-3,-24){\circle*{6}}
 \put(-6,-3){\vector(0,-1){14}}
 \put(0,0){\vector(1,0){16}}
 \put(-6,0){\circle{6}}
 \put(0,6){\circle*{2}}
     \put(-10,8){\footnotesize $n$}
     \put(-25,-20){\footnotesize $\sigma \rho$}
     \put(20,-18){\makebox(8,6){\footnotesize
      $\epsilon (-)^{n_{\mu}}\mu$}}
     \put(50,15){\makebox(8,6){\footnotesize
      $\delta\epsilon (-)^{n_{\nu}}\nu$}}
}
\end{picture}
\hspace{1cm}
-
\hspace{2.2cm}
\begin{picture}(50,50)
 \allinethickness{0.4mm}
\put(0,0){
 \put(0,0){\path(-3,6)(-24,6)(-24,0)(0,0)}
 \put(0,0){\path(6,0)(30,0)(30,30)(0,30)(0,9)}
 \thicklines
 \put(-24,3){\circle*{6}}
 \put(-3,6){\vector(-1,0){14}}
 \put(6,0){\vector(1,0){16}}
 \put(0,6){\circle{6}}
 \put(0,0){\circle*{2}}
 \put(6,0){\circle*{2}}
     \put(-6,-10){\footnotesize $n$}
     \put(-18,10){\footnotesize $\sigma \rho$}
     \put(25,-15){\makebox(8,6){\footnotesize
      $\epsilon (-)^{n_{\mu}}\mu$}}
     \put(50,15){\makebox(8,6){\footnotesize
      $\delta\epsilon (-)^{n_{\nu}}\nu$}}
}
\end{picture}
\hspace{1.5cm}
\biggr)
\Biggr).
\label{f-linear fig}
\end{eqnarray}
\\
\noindent
%%%%%%%%%%%%%%%%%%%%%%%%%%%%%%%%%%%%%%%%%%%%%%%%%%%%%%%%%%%%%%%%%%%%%%
%%%%%%%%%%%%%%%%%%%%%%%%%%%%%%%%%%%%%%%%%%%%%%%%%%%%%%%%%%%%%%%%%%%%%%
{}From eq.~\bref{f-linear fig}, we may extract a relation between the
transformation parameters.  In doing that, remember to follow the arrows
in writing down equations from figures.  Finally we obtain the
relation,
%%%%%%%%%%%%%%%%%%%%%%%%%%%%%%%%%%%%%%%%%%%%%%%%%%%%%%%%%%%%%%%%%%%%%%
%%%%%%%%%%%%%%%%%%%%%%%%%%%%%%%%%%%%%%%%%%%%%%%%%%%%%%%%%%%%%%%%%%%%%%
\begin{eqnarray}
\frac{1}{2}b_{\pm \rho}(n)C_{n,\mu\nu}^{\delta, \epsilon}-
\frac{\beta_{\delta}}{2}
\Bigl(
\alpha^{\pm \rho}_{n,\epsilon(-)^{n_{\mu}}\mu}-
\alpha^{\pm \rho}_{n,\delta \epsilon(-)^{n_{\nu}}\nu}
\Bigr)=0.
\label{f-linear invariance}
\end{eqnarray}
%%%%%%%%%%%%%%%%%%%%%%%%%%%%%%%%%%%%%%%%%%%%%%%%%%%%%%%%%%%%%%%%%%%%%%
%%%%%%%%%%%%%%%%%%%%%%%%%%%%%%%%%%%%%%%%%%%%%%%%%%%%%%%%%%%%%%%%%%%%%%
The first term of eq.~\bref{f-linear invariance} is from the
transformation of $\xi_n$ in $S_f$.  The other two terms are from
$\delta_U S_g^{\delta}$ : the second (third) term is obtained by
transforming the link variable $U_{n,\epsilon(-)^{n_{\mu}}\mu}$
(~$U_{n,\delta \epsilon(-)^{n_{\nu}}\nu}$~).  Note that, contrary to
eq.~\bref{fermion cubed condition}, eq.~\bref{f-linear invariance} is
$\delta$-dependent.  It is also important to remember that $\pm \rho$
can take any direction, independently $\mu$ and $\nu$.

%%%%%%%%%%%%%%%%%%%%%%%%%%%%%%%%%%%%%%%%%%%%%%%%%%%%%%%%%%%%%%%%%%%%%
%%%%%%%%%%%%%%%%%%%%%%%%%%%%%%%%%%%%%%%%%%%%%%%%%%%%%%%%%%%%%%%%%%%%%
\subsection{Invariance of the cell + pipe mixed system}
%%%%%%%%%%%%%%%%%%%%%%%%%%%%%%%%%%%%%%%%%%%%%%%%%%%%%%%%%%%%%%%%%%%%%
%%%%%%%%%%%%%%%%%%%%%%%%%%%%%%%%%%%%%%%%%%%%%%%%%%%%%%%%%%%%%%%%%%%%%

Up to now, we have studied the cell or pipe model separately.  Here
we would like to introduce their mixed system and derive the conditions
for its preSUSY invariance.

The actions for the cell and pipe models differ in the allowed
plaquettes, while the fermion action is independent of $\delta$.  As for
the preSUSY transformation, the model dependence appears only in
eq.~\bref{fermion tf}.  These observations are made explicit in the
following equations,
\begin{eqnarray}
S = S_f + S_g^{cell} + S_g^{pipe},~~~
\delta = \delta_U + \delta_{\xi}^{cell} + \delta_{\xi}^{pipe}.
\end{eqnarray}
Here the model dependence is written explicitly, rather than indicating
it by the sign factor $\delta$.  The gauge action consists of two parts,
each of them is multiplied by the overall coefficient $\beta_{\delta}$.
When two $\beta$ coincide, $\beta_{+1}=\beta_{-1}$ the gauge action is
an ordinary plaquette action.

In the transformation of the action
\begin{eqnarray}
\delta S = \delta_U S_f + \delta_U S_g^{cell} + \delta_U S_g^{pipe}
+ \delta_{\xi}^{cell} S_f + \delta_{\xi}^{pipe} S_f=0,
\end{eqnarray}
three different types of diagrams vanish independently:  
\begin{eqnarray}
\delta_U S_f &=&0,\nonumber\\
\delta_U S_g^{cell} &+& \delta_{\xi}^{cell} S_f =0,\label{pipe conditions}\\
\delta_U S_g^{pipe} &+& \delta_{\xi}^{pipe} S_f =0.\nonumber
\end{eqnarray}
Only the first equation contains fermion cubic terms.  In the other two
equations, different set of plaquettes appear.  So it vanishes
independently.

Therefore for the invariance of the action, the transformation
parameters must satisfy eqs. \bref{fermion cubed condition} and
\bref{f-linear invariance}.  In particular, the latter must hold both for
$\delta=\pm 1$.  These conditions will be solved in the next section.

A comment is in order for the number of transformation parameters in the
case of the mixed model.  $\delta_U$ is the same both for the cell or
pipe model.  So even for the mixed model, we have the same number of
$\alpha$-parameters.  The $\delta$ dependence shows up in the fermion
transformation.  The number of $C$-parameters is doubled for the mixed
model, accordingly.  Though we solved eqs. \bref{fermion cubed
condition} and \bref{f-linear invariance} for the cell model, it is not
so obvious that we could do the same for the mixed model.  This will be
discussed in the following section.

%%%%%%%%%%%%%%%%%%%%%%%%%%%%%%%%%%%%%%%%%%%%%%%%%%%%%%%%%%%%%%%%%%%%%
\section{Solving conditions for the invariance}
%%%%%%%%%%%%%%%%%%%%%%%%%%%%%%%%%%%%%%%%%%%%%%%%%%%%%%%%%%%%%%%%%%%%%

We obtained the conditions for the action invariance \bref{fermion cubed
condition} and \bref{f-linear invariance} listed below:
%%%%%%%%%%%%%%%%%%%%%%%%
\begin{eqnarray}
b_{\sigma \mu}(n) \alpha_{n,\sigma \mu}^{\sigma^{\prime} \rho}
&+& b_{\sigma^{\prime} \rho}(n)
\alpha_{n,\sigma^{\prime} \rho}^{\sigma \mu} =0,
\label{fermion cubed condition 2}\\
b_{\sigma \rho}(n)C_{n,\mu\nu}^{\delta, \epsilon}&-&
{\beta_{\delta}}
\Bigl(
\alpha^{\sigma \rho}_{n,\epsilon(-)^{n_{\mu}}\mu}-
\alpha^{\sigma \rho}_{n,\delta \epsilon(-)^{n_{\nu}}\nu}
\Bigr)=0~~~~~(\mu<\nu).
\label{f-linear invariance 2}
\end{eqnarray}
%%%%%%%%%%%%%%%%%%%%%%%%
The restriction $\mu<\nu$ is due to the definition of
$C_{n,\mu\nu}^{\delta, \epsilon}$.  Because of this, we need some care
in writing equations.  In this section, when we have a similar
restriction, we write it explicitly as eq.~\bref{f-linear invariance 2}
to avoid confusion.

The first equation tells us that
$b_{\sigma \mu}(n) \alpha_{n,\sigma \mu}^{\sigma^ {\prime} \rho}$ is
antisymmetric under the exchange of $\sigma \mu \leftrightarrow
\sigma^{\prime} \rho$.  From the relation, we may obtain, eg,
$\alpha_{n,\sigma \mu}^{\sigma \mu} =0$ or
$\alpha_{n,\mu}^{-\mu}-\alpha_{n,-\mu}^{\mu}=0$.  By solving
eq.~\bref{f-linear invariance 2} for one of the $\alpha$'s, we find that
the lower index of the $\alpha$'s can be changed by adding the
$C$-parameter.

For the cell model, we solved eqs. \bref{fermion cubed condition 2} and
\bref{f-linear invariance 2} with $\delta=+1$ and find $(2D-1)$
independent parameters at each site\cite{1st}.  At this moment it is not
clear whether we can do the same for the pipe model or the mixed model.
For the pipe model, we consider \bref{fermion cubed condition 2} and
\bref{f-linear invariance 2} with $\delta=-1$, while for the mixed model
we must solve three conditions, \bref{fermion cubed condition 2} and
\bref{f-linear invariance 2} with $\delta=\pm 1$, corresponding to
three equations given in \bref{pipe conditions}.  Thus one may consider
that we solve different sets of equations for the pipe and mixed models.
However we actually can show that effectively the same set of equations
are to be solved for both of the mixed and the pipe models.  We first
explain this observation.

Consider the conditions for pipe model.  
Eqs.~\bref{fermion cubed condition 2} and \bref{f-linear
invariance 2} hold at each site independently.  So let us study the
conditions at the origin $n=0$ for simplicity.  This makes us easy to
understand the following discussion without being too much bothered by
various sign factors.  In appendix A, we will solve the conditions at a generic
site.
As for $b_{\mu}(n)$, we use the expression to be given in the next
section, which produces a Majorana fermion in the continuum limit.  At
the origin, it takes the values as $b_{\pm \mu}(n=0)=\pm 1$.  Thus
eqs. \bref{fermion cubed condition 2} and \bref{f-linear invariance 2}
for $\delta=-1$ are written as
\begin{eqnarray}
\sigma \alpha_{~~\sigma \mu}^{{\sigma}^{\prime}\rho}
&+&{\sigma}^{\prime} \alpha_{~~{\sigma}^{\prime} \rho}^{\sigma \mu}=0,
\label{antisym}\\
\bar{C}_{\mu \nu}^{(\epsilon)} &-& \sigma {\beta_p} 
\Bigl(\alpha_{~~\epsilon \mu}^{\sigma \rho}
-\alpha_{~- \epsilon \nu}^{\sigma \rho}\Bigr)=0~~~~(\mu~<~\nu).
\label{C and alpha}
\end{eqnarray}
Here we dropped the site index and used the notations $\bar{C}$ and
${\beta_p}$ to represent the quantities for $\delta=-1$.  

It is easy to derive 
\begin{eqnarray}
\bar{C}^{(\epsilon)}_{\lambda\mu} + \bar{C}^{(-\epsilon)}_{\mu\nu}
=\sigma {\beta_p}\Bigl( \alpha_{~\epsilon \lambda}^{\sigma \rho}
-\alpha_{~\epsilon \nu}^{\sigma \rho} \Bigr)~~~~~(\lambda<\mu < \nu). 
\label{C sum}
\end{eqnarray}
Though the lhs depends on $\epsilon$, $\mu$, $\nu$ and $\lambda$, the
rhs does not depend on the index $\mu$.  On top of that, the combination
on the rhs of eq.~\bref{C sum} appears in eq.~\bref{f-linear invariance
2} written for $\delta=+1$ and $n=0$,
\begin{eqnarray}
C_{\lambda\nu}^{(\epsilon)}
\equiv \sigma {\beta_c} \Bigl(\alpha_{~\epsilon \lambda}^{\sigma \rho}
-\alpha_{~\epsilon \nu}^{\sigma \rho}\Bigr)~~~~~~~~~(\lambda<\nu).
\label{def C}  
\end{eqnarray}
Thus, for the pipe model, we solve three relations \bref{antisym}, \bref{C and alpha} and
\bref{def C}; the last relation is regarded as the definitions of
$C_{\lambda\nu}^{(\epsilon)}$ and $\beta_c$.  This is the same set of
equations to be solved for the mixed model.  Therefore, by solving these
three relations, we can consider the pipe and mixed models at once.

Note that $C$ in eq.~\bref{def C} enjoys the property
$C_{\lambda\mu}^{(\epsilon)}+C_{\mu\nu}^{(\epsilon)}-C_{\lambda\nu}^{(\epsilon)}=0~~~~~(\lambda~<~\mu~<~\nu)$
which may be solved as\cite{1st}
\begin{eqnarray} 
C_{\mu\nu}^{(\epsilon)}=C_{1\nu}^{(\epsilon)}-C_{1\mu}^{(\epsilon)}
~~~~~~~~~(1<\mu < \nu).
\label{sol for C}
\end{eqnarray}

{}From eq.~\bref{antisym}, we learn that diagonal elements of the
$\alpha$-parameter vanish.  Using this fact, we may derive the
relations between $\alpha$-parameters and $C$-parameters from
eqs.~\bref{C and alpha} and \bref{def C},
\begin{eqnarray}
C_{\mu\nu}^{(\epsilon)} &=& \epsilon {\beta_c} \alpha_{~\epsilon \mu}^{\epsilon \nu}=
-\epsilon {\beta_c} \alpha_{~\epsilon \nu}^{\epsilon \mu}~~~~~~~~~~~~(\mu<\nu),
\label{relation C and alpha 1}\\
{\bar C}_{\mu\nu}^{(\epsilon)} &=& -\epsilon {\beta_p} \alpha_{~\epsilon \mu}^{-\epsilon \nu}=
-\epsilon {\beta_p} \alpha_{~-\epsilon \nu}^{\epsilon \mu}~~~~~~~~~(\mu<\nu).
\label{relation C and alpha 2}  
\end{eqnarray}
{}For example, eq.~\bref{relation C and alpha 2} is obtained from
eq.~\bref{C and alpha} by taking $\sigma \rho = -\epsilon \nu$ or
$\epsilon \mu$.

As transformation parameters, $C_{\mu\nu}^{(\epsilon)}$ and ${\bar
C}_{\mu\nu}^{(\epsilon)}$ are associated with the plaquettes for the
cell and pipe models, respectively.  Through eqs.~\bref{relation C and
alpha 1} and \bref{relation C and alpha 2}, we may relate these
$\alpha$-parameters to the cell and pipe plaquettes.  Note that
eq.~\bref{C and alpha} contains another type of $\alpha$-parameters,
$\alpha_{~\mu}^{-\mu} = \alpha_{~-\mu}^{\mu}$, which do not appear in
eqs.~\bref{relation C and alpha 1} and \bref{relation C and alpha 2}.
Thus there are three types of $\alpha$-parameters.

Next we show that all the parameters may be written in terms of
$\alpha_1^{-1}$ and ${C}^{(\epsilon)}_{1\mu}$ (or
$\bar{C}^{(\epsilon)}_{1\mu}$).

By using eqs.~\bref{sol for C} and \bref{relation C and alpha 1}, the
$\alpha$-parameters associated with the cell model may be written as
\begin{eqnarray}
\alpha_{~\epsilon \mu}^{\epsilon \nu}=
-\alpha_{~\epsilon \nu}^{\epsilon \mu}
=\frac{\epsilon}{\beta_c}\Bigl( C^{(\epsilon)}_{1 \nu}-C^{(\epsilon)}_{1 \mu}\Bigr)
~~~~~~~~~~~~(\mu<\nu).
\label{alpha 1}
\end{eqnarray}

Next consider the $\alpha$-parameters associated with the pipe model,
$\alpha_{~\epsilon \mu}^{-\epsilon \nu} = \alpha_{~-\epsilon
\nu}^{\epsilon \mu}$.  Let us write the relation to shift the lower
index of an $\alpha$-parameter, which is obtained from \bref{def C} by
the replacements, $\lambda \rightarrow 1$ and $\nu \rightarrow \mu$,
\begin{eqnarray}
\alpha_{~\epsilon \mu}^{\sigma \rho} 
=  \alpha_{~\epsilon 1}^{\sigma \rho} - \frac{\sigma}{\beta_c}C^{(\epsilon)}_{1\mu}
~~~~~~~~(\mu>1).
\label{shift index}
\end{eqnarray} 
Using eqs. \bref{antisym} and \bref{shift index}, we find that
\begin{eqnarray}
\alpha_{~\epsilon \mu}^{-\epsilon \nu} 
&=&  \alpha_{~\epsilon 1}^{-\epsilon \nu} + \frac{\epsilon}{\beta_c}C^{(\epsilon)}_{1\mu}
= \alpha^{\epsilon 1}_{~-\epsilon \nu} + \frac{\epsilon}{\beta_c}C^{(\epsilon)}_{1\mu}
= \alpha^{~\epsilon 1}_{-\epsilon 1} 
+ \frac{\epsilon}{\beta_c}\Bigl(C^{(\epsilon)}_{1\mu}-C^{(-\epsilon)}_{1\nu}\Bigr)
\nonumber\\ 
&=& \alpha_{~1}^{-1} 
+ \frac{\epsilon}{\beta_c}\Bigl(C^{(\epsilon)}_{1\mu}-C^{(-\epsilon)}_{1\nu}\Bigr)~~~~~~~~~~~~~~(1<\mu,\nu,~~~\mu \ne \nu).
\label{alpha as C}
\end{eqnarray}
It would be instructive to explain the above calculation.  In the
first equality, we shifted the lower index of $\alpha$ by using
eq.~\bref{shift index} with $\sigma \rightarrow -\epsilon$ and $\rho
\rightarrow \nu$.  In the second equality, we exchanged the upper and
lower indices by using eqs.~\bref{antisym}.  We again shifted the (new)
lower index in the next expression.

Eq.~\bref{alpha as C} with \bref{relation C and alpha 2} gives us the
expression of ${\bar C}_{\mu\nu}^{(\epsilon)}$ in terms of
$\alpha_1^{-1}$ and ${C}^{(\epsilon)}_{1\mu}$,
\begin{eqnarray}
\bar{C}_{\mu \nu}^{(\epsilon)}=-\epsilon \beta_p \alpha_{~1}^{-1} 
- \frac{\beta_p}{\beta_c}\Bigl(C^{(\epsilon)}_{1\mu}-C^{(-\epsilon)}_{1\nu}\Bigr)~~~~~~~(\mu<\nu).
\label{expression of bar C}
\end{eqnarray}

We obtain the expression for $\alpha_{-\mu}^{\mu}$ as follows,
\begin{eqnarray}
\alpha_{-\mu}^{\mu}&=&\alpha_{~\epsilon \mu}^{-\epsilon \mu} 
=\alpha_{~~\epsilon 1}^{-\epsilon \mu}  + \frac{\epsilon}{\beta_c}C^{(\epsilon)}_{1\mu}=\alpha_{-\epsilon 1}^{\epsilon 1}  - \frac{\epsilon}{\beta_c}\Bigl(C^{(-\epsilon)}_{1\mu}-C^{(\epsilon)}_{1\mu}\Bigr)\nonumber\\
&=&\alpha_{~-1}^{1}  + \frac{1}{\beta_c}\Bigl(C^{(+)}_{1\mu}-C^{(-)}_{1\mu}\Bigr)
~~~~~~~~(\mu>1).
\label{alpha 3}
\end{eqnarray} 
Here, in the first equality, use was made of eq.~\bref{shift index} with
$\sigma \rightarrow -\epsilon$ and $\rho \rightarrow \mu$.  

In summary, we have derived expressions of all the parameters by
independent ones, given in eqs.~\bref{sol for C}, \bref{alpha 1},
\bref{alpha as C}, \bref{expression of bar C} and \bref{alpha 3}.
It is easy to confirm the validity of our results by substituting them
back to eqs.~\bref{antisym}, and \bref{C and alpha} and \bref{def C}.

Studying relations on the parameters at the origin, we have shown
explicitly that all the parameters can be expressed with
$\alpha_{~~1}^{-1}$ and ${C}^{(\epsilon)}_{1\mu}~~(\mu>1)$.  However, it
must be obvious that our discussion may be applicable to other sites,
that is to be done in appendix A.  There, we find 2D-1 independent
parameters at each site, $\alpha_{n,1}^{-1}$ and
${C}^{(\epsilon)}_{n,1\mu}~~(\mu>1)$, the same number as the cell model.
Of course, we may choose a different set of independent parameters.  In
particular, we can take $\bar{C}_{n,1\mu}^{(\epsilon)}$ instead of
${C}_{n,1\mu}^{(\epsilon)}$ (cf.~eq.~\bref{C and Cbar}).  Leaving the
details to appendix A, we quote the expressions of parameters in
terms of independent parameters,
\begin{eqnarray}
C_{n,\mu\nu}^{(\epsilon)} &=& C_{n,1\nu}^{(\epsilon)}
-C_{n,1\mu}^{(\epsilon)}
=\frac{\beta_c}{\beta_p}\Bigl(
\bar{C}_{n,1\nu}^{(-\epsilon)}
-\bar{C}_{n,1\mu}^{(-\epsilon)}
\Bigr)~~~~~(\mu<\nu),
\label{C mu nu}\\
\bar{C}_{n,\mu\nu}^{(\epsilon)} 
&=&
\frac{\beta_p}{\beta_c}\Bigl({C}_{n,1\nu}^{(-\epsilon)}
-{C}_{n,1\mu}^{(\epsilon)}\Bigr) -\epsilon {\beta_p} (-)^{n_1} \alpha_{n,1}^{-1}
\nonumber\\
&=& \bar{C}_{n,1\nu}^{(\epsilon)}
-\bar{C}_{n,1\mu}^{(-\epsilon)} +\epsilon {\beta_p} (-)^{n_1} \alpha_{n,1}^{-1}~~~~~(\mu<\nu).
\label{relation of C}
\end{eqnarray}
The $\alpha$-parameters may be written as
\begin{eqnarray}
&{}&\alpha_{n,\delta \epsilon (-)^{n_{\nu}} \nu}^{\epsilon (-)^{n_{\mu}} \mu} =
- \frac{\epsilon (-)^{n_{\mu}}}{\beta_{\delta}} b_{\mu}(n)
C_{n,\mu\nu}^{\delta, \epsilon}~~~~~(\mu<\nu), 
\label{alpha C delta}\\
&{}&\alpha_{n, \nu}^{-\nu} =
(-)^{n_{\nu}+n_1}b_{\nu}(n) \alpha_{n,1}^{-1}
+\frac{\epsilon (-)^{n_{\nu}}}{\beta_c} b_{\nu}(n)
\Bigl(C_{n,1\nu}^{(\epsilon)}-C_{n,1\nu}^{(-\epsilon)}\Bigr)\nonumber\\
&{}&~~~~~=
- (-)^{n_{\nu}+n_1}b_{\nu}(n) \alpha_{n,1}^{-1}
-\frac{\epsilon (-)^{n_{\nu}}}{\beta_p} b_{\nu}(n)
\Bigl(\bar{C}_{n,1\nu}^{(\epsilon)}-\bar{C}_{n,1\nu}^{(-\epsilon)}\Bigr)~~~(\nu>1)
\label{alpha nu nu}
\end{eqnarray}
where $\beta_c \equiv \beta_{\delta=+1}$.

We conclude that above equations solve the conditions for any model,
ie, the cell, the pipe and even the mixed model.

%%%%%%%%%%%%%%%%%%%%%%%%%%%%%%%%%%%%%%%%%%%%%%%%%%%%%%%%%%%%%%%%%%%%%
%%%%%%%%%%%%%%%%%%%%%%%%%%%%%%%%%%%%%%%%%%%%%%%%%%%%%%%%%%%%%%%%%%%%%
%%%%%%%%%%%%%%%%%%%%%%%%%%%%%%%%%%%%%%%%%%%%%%%%%%%%%%%%%%%%%%%%%%%%%
\section{The Majorana staggered fermion}
%%%%%%%%%%%%%%%%%%%%%%%%%%%%%%%%%%%%%%%%%%%%%%%%%%%%%%%%%%%%%%%%%%%%%%
%%%%%%%%%%%%%%%%%%%%%%%%%%%%%%%%%%%%%%%%%%%%%%%%%%%%%%%%%%%%%%%%%%%%%%

Here we consider the reconstruction problem and show that our fermion
action in eq.~\bref{fermion action} produces a Majorana fermion in a
naive continuum limit.  We briefly summarize our results and describe
the details in appendix C.

%%%%%%%%%%%%%%%%%%%%%%%%%%%%%%%%%%%%%%%%%%%%

We consider the following action,
\begin{eqnarray}
S_f = \sum_n \eta_{\mu}(n) \xi_n \xi_{n+\hmu}.
\label{lattice Sf}
\end{eqnarray}
The coefficient $b_{\mu}(n)$ in eq.~\bref{fermion action} is chosen to
be $\eta_{\mu}(n) \equiv (-)^{n_1+ \cdots n_{\mu-1}}$.  As explained in
appendix C, this coefficient $\eta_{\mu}(n)$ is naturally obtained when
we start from the naively discretized Majorana fermion.  The fermion
variable $\xi_n$ satisfies the reality condition
\begin{eqnarray}
\xi_n^{\dagger}=(-)^{|n|}\xi_n
\label{xi reality}
\end{eqnarray}
so that the action in eq.~\bref{lattice Sf} is real.  The condition
given in eq.~\bref{xi reality} is also discussed in appendix C.

For the reconstruction, we make the identification $\xi_r(N) \equiv
\xi_{2N+r}$, where $2N$ is the reference coordinate of a hypercube and
$r$ is the relative coordinate in the hypercube.  In rewriting the action 
\bref{lattice Sf}, we obtain 
\begin{eqnarray}
S_f &=& \frac{1}{2}\sum_n \eta_{\mu}(n) \xi_n (\xi_{n+\hmu}-\xi_{n-\hmu})
\nonumber\\
&=& \frac{1}{2} \sum_{N,r,\mu>0}\eta_{\mu}(r) \xi_r(N){\hat \partial}_{\mu}
\Bigl[ \xi_{r+\hmu}(N) + \xi_{r-\hmu}(N) \Bigr] + \cdots.
\label{lattice Sf2}
\end{eqnarray}
The higher order terms in the naive continuum limit are denoted by
dots.  The difference ${\hat \partial}_{\mu}$ is defined as 
\begin{eqnarray}
{\hat \partial}_{\mu}  \xi_{r^{\prime}}(N) &\equiv&
\frac{1}{2} \Bigl( 
\xi_{r^{\prime}}(N+{\hat \mu}) -
\xi_{r^{\prime}}(N-{\hat \mu}) \Bigr).
\nonumber
\end{eqnarray}

Note that, in the reconstruction of the Dirac fermion, we have ${\bar
\xi}_r(N)$ instead of ${\xi}_r(N)$ in the second line of eq. \bref{lattice
Sf2}. In that case, we reach the ordinary action 
\begin{eqnarray}
S_{\rm Dirac} &=& \sum_{N,\mu>0} {\bar {\hat \psi}}_{i \alpha}(N) \gamma_{\alpha \beta}{\hat \partial}_{\mu}
{\hat \psi}_{\beta i}(N)+ \cdots,
\label{reconstructed Sf}
\end{eqnarray}
by using the relations,
\begin{eqnarray}
\xi_r(N) &=& \frac{1}{{\cal N}_0} {\rm tr} \Bigl[
V_r^{\dagger} {\hat \psi}(N)\Bigr],\nonumber\\
{\bar \xi}_r(N) &=& \frac{1}{{\cal N}_0} {\rm tr} \Bigl[
{\bar {\hat \psi}}(N) V_r\Bigr],
\label{xi and psi}
\end{eqnarray}
and the completeness of $V_r \equiv \gamma_1^{r_1}\gamma_2^{r_2} \cdots
\gamma_D^{r_D}$.  ${{\cal N}_0}$ is a normalization coefficient.  In
eq. \bref{reconstructed Sf}, the spinor and flavor indices are written
as $\alpha$ and $i$, respectively.

In the Majorana staggered fermion case, $\xi_r(N)$ must represent two
expressions on the rhs of eq. \bref{xi and psi}.  This condition may be
realized as the Majorana condition for ${\hat \psi}$ 
\begin{eqnarray}
{\hat \psi} = C {\bar {\hat \psi}}^T C^{-1}.
\label{Maj cond on psi}
\end{eqnarray}
Note that $C^{-1}$ acts on the ``flavor'' index.  From
Ref. \cite{KugoTownsend}, we find the dimensions to realize the
Majorana condition \bref{Maj cond on psi}: $D=1,2,8~~({\rm mod}~8)$ for
Majorana fermions; $D=4,5,6~~({\rm mod}~8)$ for Usp Majorana
fermions. The (pseudo-)Majorana fermions in $D=3,7,11$ are not realized
this way.

The reality of $\xi_n$ is translated into the condition on ${\hat \psi}$, 
\begin{eqnarray}
{\bar {\hat \psi}}= \gamma_{D+1}^{\dagger} {\hat \psi}^{\dagger} \gamma_{D+1},
\label{psi bar and psi dagger 1}
\end{eqnarray}
where $\gamma_{D+1}$ is a hermitian matrix
\begin{eqnarray}
\gamma_{D+1} \equiv i^{D/2} \gamma_1 \cdots \gamma_D.
\label{gamma D+1 1}
\end{eqnarray}
Since $\gamma_{D+1}$ can be used as eq.~\bref{psi bar and psi dagger 1}
only in the even dimensional spaces, the dimensions listed in the last
paragraph must be restricted accordingly.  The odd dimensional spaces
must be considered separately; the subject will not be discussed in the
present paper.

%%%%%%%%%%%%%%%%%%%%%%%%%%%%%%%%%%%%%%%%%%%%%%%%%%%%%%%%%%%%%%%%%%%%%%
%%%%%%%%%%%%%%%%%%%%%%%%%%%%%%%%%%%%%%%%%%%%%%%%%%%%%%%%%%%%%%%%%%%%%%
%%%%%%%%%%%%%%%%%%%%%%%%%%%%%%%%%%%%%%%%%%%%%%%%%%%%%%%%%%%%%%%%%%%%%%
\section{On the continuum limit}
%%%%%%%%%%%%%%%%%%%%%%%%%%%%%%%%%%%%%%%%%%%%%%%%%%%%%%%%%%%%%%%%%%%%%%
%%%%%%%%%%%%%%%%%%%%%%%%%%%%%%%%%%%%%%%%%%%%%%%%%%%%%%%%%%%%%%%%%%%%%%
%%%%%%%%%%%%%%%%%%%%%%%%%%%%%%%%%%%%%%%%%%%%%%%%%%%%%%%%%%%%%%%%%%%%%%

When we introduced our preSUSY transformation, we first assumed the form
of the fermion transformation given in eq.~\bref{fermion tf}.  The form was
chosen with the expectation that it would produce the continuum SUSY
transformation, $\delta \psi \sim \gamma_{\mu\nu} F_{\mu\nu}\epsilon$.
Without fully studying the spinor structure, it is not clear whether
this is achieved or not.  However, as a step toward it, we would like to
see its form in the naive continuum limit.  This is to be discussed in
the next subsection.

The pattern we put the plaquettes may look peculiar.  Since we expect an
appropriate continuum limit, we confirm, in the second subsection, that
our models satisfy reasonable requirements for such a limit.  That is,
we show that our models have translational and rotational invariances,
and satisfy the condition of the reflection positivity\cite{OS2}.  

%%%%%%%%%%%%%%%%%%%%%%%%%%%%%%%%%%%%%%%%%%%%%%%%%%%%%%%%%%%%%%%%%%%%%%
\subsection{Continuum limit of the fermion transformation}
%%%%%%%%%%%%%%%%%%%%%%%%%%%%%%%%%%%%%%%%%%%%%%%%%%%%%%%%%%%%%%%%%%%%%%

Here we will show that eq.~\bref{fermion tf} becomes
\begin{eqnarray}
\delta \xi_n \sim \Bigl(1-\frac{\beta_p}{\beta_c}\Bigr)
2 i a^{D/2} g \sum_{0<\mu<\nu} (-)^{n_{\mu}+n_{\nu}} 
\Bigl(C_{n,\mu\nu}^{(+)}+C_{n,\mu\nu}^{(-)}\Bigr)F_{n,\mu\nu}
\label{naive limit of fermion tf}
\end{eqnarray}
in the naive continuum limit.  On the rhs, there follows the terms of
the higher order in $a$.  So, on the dimensional ground, we would like
to keep the rhs of \bref{naive limit of fermion tf}, the term with the
field strength.  If the rhs vanishes, the transformation is higher order
in $a$ and vanishes in the continuum limit.  Therefore two couplings
$\beta_c$ and ${\beta_p}$ must be different.  This implies the
importance of the Ichimatsu pattern.

Now we derive eq.~\bref{naive limit of fermion tf}.  A plaquette is
related to the field strength as
\begin{eqnarray}
U_{n,\mu\nu}^{\delta,\epsilon} \sim 1+i~a^{D/2} g F_{n,\mu\nu}^{\delta, \epsilon}.
\label{cont plaquette}
\end{eqnarray}
Since the direction of the plaquettes are defined as indicated
in eq.~\bref{gauge action} or Fig.~3, we easily find the relation of the
field strength $F_{n,\mu\nu}^{\delta, \epsilon}$ to the
ordinary $F_{n,\mu\nu}$ as
\begin{eqnarray}
F_{n,\mu\nu}^{\delta, \epsilon} = \delta (-)^{n_{\mu}+n_{\nu}} F_{n,\mu\nu}.
\label{field strength relation}
\end{eqnarray} 
Using \bref{cont plaquette} and \bref{field strength relation} on the
rhs of eq.~\bref{fermion tf}, we find
\begin{eqnarray*}
\delta \xi_n &\sim& 2 i a^{D/2} g \sum_{\epsilon, \delta}\sum_{0<\mu<\nu} C_{n,\mu\nu}^{\delta, \epsilon} \delta (-)^{n_{\mu}+n_{\nu}} F_{n,\mu\nu}\nonumber\\
&=& 2ia^{D/2} g \sum_{0<\mu<\nu} (-)^{n_{\mu}+n_{\nu}} 
\Bigl(C_{n,\mu\nu}^{(+)}+C_{n,\mu\nu}^{(-)}
-\bar{C}_{n,\mu\nu}^{(+)}-\bar{C}_{n,\mu\nu}^{(-)}\Bigr)
F_{n,\mu\nu}.
\end{eqnarray*}
Now, note that the relation,
\begin{eqnarray*}
\bar{C}_{n,\mu\nu}^{(+)}+\bar{C}_{n,\mu\nu}^{(-)} = \frac{\beta_p}{\beta_c}
\Bigl(C_{n,\mu\nu}^{(+)}+C_{n,\mu\nu}^{(-)}\Bigr),
\end{eqnarray*}
may be obtained from eq.~\bref{relation of C}.  Thus we have the
transformation \bref{naive limit of fermion tf}.

Before closing this subsection, it would be worth pointing out that the
gauge coupling $g$ is related to $\beta_c$ and $\beta_p$ as
\begin{eqnarray}
\frac{1}{g^2} = \beta_c + \beta_p.
\label{g and beta}
\end{eqnarray}
In the naive continuum limit, the link field may be written as $U_{n,\mu}
= {\exp}(i~a^{\frac{D-2}{2}}~g~A_{n,\mu})$.  Accordingly, the action becomes the sum of
terms such as, $a^D \beta_c g^2 F^2_{\mu\nu}+ a^D \beta_p g^2 F^2_{\mu\nu}$.
Thus the relation \bref{g and beta} follows.

%%%%%%%%%%%%%%%%%%%%%%%%%%%%%%%%%%%%%%%%%%%%%%%%%%%%%%%%%%%%%%%%%%%%%%
\subsection{Requirements for a proper continuum theory}
%%%%%%%%%%%%%%%%%%%%%%%%%%%%%%%%%%%%%%%%%%%%%%%%%%%%%%%%%%%%%%%%%%%%%%

Let us see that the cell, pipe and mixed models have some properties so
that we may expect an appropriate continuum theory: they are
translational and rotational invariant, and they have positive definite
transfer matrices.  By construction, the three models are translational
invariant under the finite translation by $2a$.  The Ichimatsu pattern
is invariant under the rotation by $\pi /2$ around the center of any
plaquette.  Our three models have this invariance on any two dimensional
plane.  That is the rotational invariance of these models.

Osterwalder and Seiler\cite{OS2} gave a sufficient condition for a
model, defined on the Euclidean lattice, to have the positive definite
transfer matrix.  Let us see that our models satisfy this condition as
well.  Take any coordinate axis, say $x_D$, and regard it as an
imaginary time axis.  We consider the reflection along this $x_D$-axis,
denoted as $\theta$,
\begin{eqnarray} 
\theta n = \theta (n_1,n_2 \cdots , n_D)  = (n_1,n_2 \cdots , -n_D),
\label{reflection}
\end{eqnarray}
where $(n_1,n_2 \cdots , n_D)$ is the coordinate of a site $n$.  In this
subsection we assume that $n_i$ is a half-integer so that all the sites
are reflected under the above action.

Suppose we may define the action of $\theta$ on the fields so that the
action $S$ may be written as
\begin{eqnarray}
S= f + \theta f + \sum_i (\theta g_i) g_i,
\label{condition 1}
\end{eqnarray}
for some $f$ and $g$.  Then, the transfer matrix is positive
definite\cite{OS2}.  So it is important to find out that we can define
appropriate transformation of the fields under \bref{reflection}.

As for the link variables, we take the temporal gauge.  So it holds that
\begin{eqnarray}
U_{n,D}=\theta U_{n,D} =1.
 \end{eqnarray}
The reflection acts on the ``space''-like link variables as 
\begin{eqnarray}
\theta U_{n,k} = U^{\dagger}_{\theta n,k}
\end{eqnarray}
where $k=1 \sim D-1$.  The action of the reflection on the fermion is
\begin{eqnarray}
\theta \xi_n  = \eta_D(n) \xi_{\theta n}.  
\label{ref on fermion}
\end{eqnarray}
Following the arguments in \cite{OS2}, the
condition \bref{condition 1} on the action may be easily shown.  

It may be said that the cell, pipe and the mixed models are natural set
of models having three properties: translational and rotational
invariances, and the reflection positivity.  However it is not trivial
that these are the only models having these properties.  In appendix D,
we consider this point under the restricted situation: we assume the
Ichimatsu pattern on every two dimensional surface of the lattice. Under
this restriction, we show that our three models are the only models
having the three properties.

%%%%%%%%%%%%%%%%%%%%%%%%%%%%%%%%%%%%%%%%%%%%%%%%%%%%%%%%%%%%%%%%%%%%%%
%%%%%%%%%%%%%%%%%%%%%%%%%%%%%%%%%%%%%%%%%%%%%%%%%%%%%%%%%%%%%%%%%%%%%%
%%%%%%%%%%%%%%%%%%%%%%%%%%%%%%%%%%%%%%%%%%%%%%%%%%%%%%%%%%%%%%%%%%%%%%
\section{Summary and Discussion}
%%%%%%%%%%%%%%%%%%%%%%%%%%%%%%%%%%%%%%%%%%%%%%%%%%%%%%%%%%%%%%%%%%%%%%
%%%%%%%%%%%%%%%%%%%%%%%%%%%%%%%%%%%%%%%%%%%%%%%%%%%%%%%%%%%%%%%%%%%%%%
%%%%%%%%%%%%%%%%%%%%%%%%%%%%%%%%%%%%%%%%%%%%%%%%%%%%%%%%%%%%%%%%%%%%%%

In order to overcome an unnatural feature of the cell model, we
introduced the cell-pipe mixed model.  The mixed model may be treated
perturbatively, contrary to the cell model.  Furthermore we showed that
the exact fermionic symmetry is present in the mixed model as well.  We
also have seen that the fermionic sector may be reconstructed as a
Majorana fermion in a naive continuum limit.  Therefore we may conclude
that the staggered Majorana fermion plus the mixed gauge system has the
exact fermionic symmetry.

We still have not figured out how the preSUSY could be related to the
continuum SUSY.  In sect.~5, we looked at the naive continuum limit of
the fermion preSUSY transformation.  There we observed how the output
could contain the field strength.  Most importantly, we noticed that the
term with the field strength survive in a naive continuum limit only
when ${\beta_p}/\beta_c \ne 1$.  This observation in sect.~5 is the only
result we have in relation to the continuum SUSY.  It is quite
interesting that the connection of our preSUSY and the SUSY shows up
based on the particular pattern of plaquettes.  In other words, we may
keep this fermionic symmetry in the continuum limit only when we have
the Ichimatsu pattern.  For the ordinary lattice ($\beta_c = \beta_p$),
the rhs of eq.~\bref{naive limit of fermion tf} vanishes and the
transformation will become the higher order in $O(a)$.  This observation
could be of vital importance.

So, if the fermionic symmetry is really related to the expected SUSY,
the continuum limit are to be studied with the cell and pipe mixed
model.  In particular, we have to approach to the limit keeping the
condition ${\beta_p}/\beta_c \ne 1$.  Therefore we would like to see how the
theory with ${\beta_p}/\beta_c \ne 1$ is related to the ordinary one
with ${\beta_p}/\beta_c = 1$.  In the forthcoming
paper\cite{forthcoming}, we study the $(\beta_p, \beta_c)$ phase
structure for the pure gauge system.

There are other questions to be resolved.  Among others, the recovery of
the spinor structure and the doubling problem are crucially important
and most difficult.  In our formulation, the doubling problem could show
up as unbalanced degrees of freedom between the fermion and the gauge
field.  The situation suggests the higher $N$ SUSY in the continuum
theory.  Otherwise we have to consider some way to remove unwanted
doublers.  To understand this point, it would be necessary to
reconstruct the spinor structure including that of the transformation
parameters.

\vspace{0.5cm}
\noindent
{\bf \large Acknowledgment}\\
\noindent
{This work is supported in part by the Grants-in-Aid for
Scientific Research No. 12640258, 12640259, 13135209, 12640256, 12440060
and 13135205 from the Japan Society for the Promotion of Science.}

\appendix

%%%%%%%%%%%%%%%%%%%%%%%%%%%%%%%%%%%%%%%%%%%%%%%%%%%%%%%%%%%%%%%%%%%%%
\section{Invariance under infinitesimal transformation}
%%%%%%%%%%%%%%%%%%%%%%%%%%%%%%%%%%%%%%%%%%%%%%%%%%%%%%%%%%%%%%%%%%%%%

In sect.~3, we solved eqs. \bref{fermion cubed condition 2} and
\bref{f-linear invariance 2} for the site $n=0$.  In this appendix, we
extend our results there to a generic site $n$.

As in sect.~3 we consider the pipe model and solve the conditions
eqs. \bref{fermion cubed condition 2} and \bref{f-linear invariance 2}
for $\delta=-1$.  The same logical steps will be taken here as sect.~3.

Rewriting eq.~\bref{f-linear invariance 2} for $\delta=-1$ in the following form
%%%%%%%%%%%%%%%%%%%%%%%%
\begin{eqnarray}
{\bar{C}_{n,\mu\nu}^{(\epsilon)}}
=
\frac{\beta_p}{b_{\sigma \rho}(n)}
\Bigl(
\alpha^{\sigma \rho}_{n,\epsilon(-)^{n_{\mu}}\mu}-
\alpha^{\sigma \rho}_{n,- \epsilon(-)^{n_{\nu}}\nu}
\Bigr)~~~~~~~~(\mu<\nu),
\label{f-linear invariance 3orig}
\end{eqnarray}
%%%%%%%%%%%%%%%%%%%%%%%%
we easily find the relation for $\lambda<\mu < \nu$
%%%%%%%%%%%%%%%%%%%%%%%%
\begin{eqnarray}
{\bar{C}^{(\epsilon)}}_{n,\lambda\mu} + {\bar{C}^{(-\epsilon)}}_{n,\mu\nu}
= \frac{\beta_p}{b_{\sigma \rho}(n)}
\Bigl(
\alpha^{\sigma \rho}_{n,\epsilon(-)^{n_{\lambda}}\lambda}-
\alpha^{\sigma \rho}_{n,\epsilon(-)^{n_{\nu}}\nu}
\Bigr) \equiv \frac{\beta_p}{\beta_c} C_{n,\lambda\nu}^{(\epsilon)},
\nonumber
\end{eqnarray}
%%%%%%%%%%%%%%%%%%%%%%%%
Here we have defined the quantity for $\lambda<\nu$
%%%%%%%%%%%%%%%%%%%%%%%%
\begin{eqnarray}
C_{n,\lambda\nu}^{(\epsilon)} \equiv
\frac{\beta_c}{b_{\sigma \rho}(n)}
\Bigl(
\alpha^{\sigma \rho}_{n,\epsilon(-)^{n_{\lambda}}\lambda}-
\alpha^{\sigma \rho}_{n,\epsilon(-)^{n_{\nu}}\nu}
\Bigr). 
\label{def C 2}
\end{eqnarray}
%%%%%%%%%%%%%%%%%%%%%%%%
There holds the relation,
$C^{(\epsilon)}_{n,\lambda\mu}+C^{(\epsilon)}_{n,\mu\nu}-C^{(\epsilon)}_{n,\lambda\nu}=0$
($\lambda<\mu<\nu$), which may be solved as
\begin{eqnarray}
C^{(\epsilon)}_{n,\mu \nu} = C^{(\epsilon)}_{n,1 \nu}-C^{(\epsilon)}_{n,1 \mu}~~~~~~~~(1<\mu<\nu).
\label{sol for Cn}
\end{eqnarray}
This is the first equality of eq.~\bref{C mu nu}.

Eq.~\bref{def C 2} is nothing but \bref{f-linear invariance 2} for $\delta=+1$.
Therefore we solve three conditions eqs. \bref{fermion cubed condition 2} and
\bref{f-linear invariance 2} for $\delta=\pm 1$ in the following discussion.

{}From the symmetry property \bref{fermion cubed condition 2}, the
diagonal elements of $\alpha$ vanish.  Thus we obtain for $\mu<\nu$,
\begin{eqnarray}
{\bar{C}_{n,\mu\nu}^{(\epsilon)}}
&=&
\frac{\beta_p}{b_{\sigma \rho}(n)}
\alpha^{\sigma \rho}_{n,\epsilon(-)^{n_{\mu}}\mu}\Bigm|_{\sigma \rho =- \epsilon(-)^{n_{\nu}}\nu}
= - \frac{\beta_p}{b_{\sigma \rho}(n)}
\alpha^{\sigma \rho}_{n,- \epsilon(-)^{n_{\nu}}\nu}\Bigm|_{\sigma \rho =\epsilon(-)^{n_{\mu}}\mu},
\label{pipe, alpha and C bar}\\
{{C}_{n,\mu\nu}^{(\epsilon)}}
&=&
\frac{\beta_c}{b_{\sigma \rho}(n)}
\alpha^{\sigma \rho}_{n,\epsilon(-)^{n_{\mu}}\mu}\Bigm|_{\sigma \rho =\epsilon(-)^{n_{\nu}}\nu}
= - \frac{\beta_c}{b_{\sigma \rho}(n)}
\alpha^{\sigma \rho}_{n, \epsilon(-)^{n_{\nu}}\nu}\Bigm|_{\sigma \rho =\epsilon(-)^{n_{\mu}}\mu},
\label{cell, alpha and C}
\end{eqnarray}
from eqs.~\bref{f-linear invariance 3orig} and \bref{def C 2}.
Using the relation 
\begin{eqnarray}
b_{\sigma \nu}(n) = \sigma b_{\nu}(n)~~~~~~(\sigma = \pm 1),
\label{sign b0}
\end{eqnarray}
which holds for $b_{\mu}(n)= \eta_{\mu}(n)$, we may further rewrite
eqs.~\bref{pipe, alpha and C bar} and \bref{cell, alpha and C}.  For
example, we find 
\begin{eqnarray}
\alpha^{-\epsilon (-)^{n_{\nu}} \nu}_{n,\epsilon (-)^{n_{\mu}} \mu} =
- \frac{\epsilon}{\beta_p} (-)^{n_{\nu}} b_{\nu}(n)
\bar{C}_{n,\mu\nu}^{(\epsilon)}~~~~~(\mu~<~\nu).
\label{alpha and C 2a}
\end{eqnarray}
Eqs.~\bref{pipe, alpha and C bar} and \bref{cell, alpha and C} are
equivalent to eq.~\bref{alpha C delta}.

Let us show that all the parameters may be written in terms of $2D-1$ of
them, ie, $\alpha_{n,1}^{-1}$ and ${C}^{(\epsilon)}_{n,1\mu}~~(\mu>1)$.  Later we
also show that ${C}^{(\epsilon)}_{n,1\mu}$ can be replaced by
$\bar{C}^{(\epsilon)}_{n,1\mu}$.

As explained for the $n=0$ case in sect.~3, there are three kinds of
$\alpha$-parameters, $\alpha_{n,\mu}^{-\mu}$, and those associated with
the cell and pipe models via the relations in eqs.~\bref{cell, alpha and
C} and \bref{pipe, alpha and C bar}, respectively.

Using eqs.~\bref{sol for Cn}, \bref{cell, alpha and C} and \bref{sign
b0}, the $\alpha$-parameter for the cell model is expressed as
\begin{eqnarray}
&{}&\epsilon(-)^{n_{\nu}}\frac{\beta_c}{b_{\nu}(n)}
\alpha^{\epsilon(-)^{n_{\nu}}\nu}_{n,\epsilon(-)^{n_{\mu}}\mu}
= - \epsilon(-)^{n_{\mu}}\frac{\beta_c}{b_{\mu}(n)}
\alpha^{\epsilon(-)^{n_{\mu}}\mu}_{n, \epsilon(-)^{n_{\nu}}\nu}\nonumber\\
&{}&~~~~~~~~~~= C^{(\epsilon)}_{n,1\nu} - C^{(\epsilon)}_{n,1\mu}~~~~~~~~~~~~~~~(1<\mu<\nu).
\label{sol alpha for cell}
\end{eqnarray}

Now consider the $\alpha$-parameter for the pipe model.  Rewriting
eq.~\bref{def C 2}, we obtain the relation
\begin{eqnarray}
\frac{1}{b_{\sigma \rho}(n)} \alpha_{n,\epsilon(-)^{n_{\mu}}\mu}^{\sigma \rho} = 
\frac{1}{b_{\sigma \rho}(n)} \alpha_{n,\epsilon(-)^{n_{1}}1}^{\sigma \rho} - \frac{1}{\beta_c}C_{n,1\mu}^{(\epsilon)}
\label{shift index1}
\end{eqnarray}
which shifts the lower index.  Using eqs.~\bref{fermion cubed condition
2} and \bref{shift index1}, we find
\begin{eqnarray}
&{}&\frac{1}{b_{\sigma \rho}(n)}~
\alpha^{\sigma \rho}_{n,\epsilon(-)^{n_{\mu}}\mu}\Bigm|_{\sigma \rho =-\epsilon(-)^{n_{\nu}}\nu}
= -
\frac{1}{b_{\sigma \rho}(n)}~
\alpha^{\sigma \rho}_{n,-\epsilon(-)^{n_{\nu}}\nu}\Bigm|_{\sigma \rho =\epsilon(-)^{n_{\mu}}\mu}\nonumber\\
&{}&~~~~~~~~~~=
\frac{1}{b_{\sigma \rho}(n)}~
\alpha^{\sigma \rho}_{n,\epsilon(-)^{n_{1}}1}\Bigm|_{\sigma \rho =-\epsilon(-)^{n_{\nu}}\nu}
- \frac{1}{\beta_c}C_{n,1\mu}^{(\epsilon)}\nonumber\\
&{}&~~~~~~~~~~=
-\frac{1}{b_{\epsilon (-)^{n_1}1}(n)}~
\alpha^{\epsilon (-)^{n_1}1}_{n,-\epsilon(-)^{n_{\nu}}\nu}
- \frac{1}{\beta_c}C_{n,1\mu}^{(\epsilon)}\nonumber\\
&{}&~~~~~~~~~~=
-\frac{1}{b_{\epsilon (-)^{n_1}1}(n)}~
\alpha^{\epsilon (-)^{n_1}1}_{n,-\epsilon(-)^{n_{1}}1}
+ \frac{1}{\beta_c}\Bigl(C_{n,1\nu}^{(-\epsilon)}-C_{n,1\mu}^{(\epsilon)}\Bigr)\nonumber\\
&{}&~~~~~~~~~~=
-\epsilon (-)^{n_1}\alpha^{-1}_{n,1}
+ \frac{1}{\beta_c}\Bigl(C_{n,1\nu}^{(-\epsilon)}-C_{n,1\mu}^{(\epsilon)}\Bigr)~~~~~~~~(1<\mu,\nu,~~~\mu\ne\nu).
\label{alpha for pipe n}
\end{eqnarray}

Let us consider $\alpha_{n,\mu}^{-\mu}$.  Again, using
eqs.~\bref{fermion cubed condition 2} and \bref{shift index1}, we obtain
\begin{eqnarray}
-\frac{\epsilon (-)^{n_{\mu}}}{b_{\mu}(n)} \alpha_{n,\mu}^{-\mu}
&=& \frac{1}{b_{\sigma \rho}(n)}
\alpha^{\sigma \rho}_{n,\epsilon(-)^{n_{\mu}}\mu}\Bigm|_{\sigma \rho =-\epsilon(-)^{n_{\mu}}\mu}
\nonumber\\
&=& \frac{1}{b_{\sigma \rho}(n)}
\alpha^{\sigma \rho}_{n,\epsilon(-)^{n_{1}}1}\Bigm|_{\sigma \rho =-\epsilon(-)^{n_{\mu}}\mu}
-\frac{1}{\beta_c}C_{n,1\mu}^{(\epsilon)}
\nonumber\\
&=&-\frac{1}{b_{\epsilon(-)^{n_{1}}1}(n)}
\alpha^{\epsilon(-)^{n_{1}}1}_{n,-\epsilon(-)^{n_{\mu}}\mu}-\frac{1}{\beta_c}C_{n,1\mu}^{(\epsilon)}
\nonumber\\
&=&-\frac{1}{b_{\epsilon(-)^{n_{1}}1}(n)}
\alpha^{\epsilon(-)^{n_{1}}1}_{n,-\epsilon(-)^{n_{1}}1}+ \frac{1}{\beta_c}
\Bigl(C_{n,1\mu}^{(-\epsilon)}-C_{n,1\mu}^{(\epsilon)}\Bigr)
\nonumber\\
&=& -\epsilon(-)^{n_1}
\alpha^{-1}_{n,1}+ \frac{1}{\beta_c}
\Bigl(C_{n,1\mu}^{(-\epsilon)}-C_{n,1\mu}^{(\epsilon)}\Bigr).
\label{alpha mu -mu}
\end{eqnarray}
This gives the first line of eq.~\bref{alpha nu nu}.
Note that eqs.~\bref{alpha for pipe n} and \bref{alpha mu -mu} can be
summarized by a single equation,
\begin{eqnarray}
\frac{(-)^{n_{\nu}}}{b_{\nu}(n)}~
\alpha^{-\epsilon(-)^{n_{\nu}}\nu}_{n,\epsilon(-)^{n_{\mu}}\mu}
=  \frac{(-)^{n_{\mu}}
}{b_{\mu}(n)}~
\alpha^{\epsilon(-)^{n_{\mu}}\mu}_{n,-\epsilon(-)^{n_{\nu}}\nu}
=
(-)^{n_1}\alpha^{-1}_{n,1}
+ \frac{\epsilon}{\beta_c}\Bigl(C_{n,1\mu}^{(\epsilon)}-C_{n,1\nu}^{(-\epsilon)}\Bigr),
\label{summary alpha for pipe n}
\end{eqnarray}
which is valid for $\mu,\nu>1$.  

Using eqs.~\bref{alpha and C 2a} and \bref{summary alpha for pipe n}, we 
find that
\begin{eqnarray}
\frac{1}{\beta_p} \bar{C}_{n,\mu\nu}^{(\epsilon)}= -\epsilon (-)^{n_1} \alpha_{n,1}^{-1}- \frac{1}{\beta_c}\Bigl( C_{n,1\mu}^{(\epsilon)}- C_{n,1\nu}^{(-\epsilon)}
\Bigr),
\label{Cbar}
\end{eqnarray}
the first equality of eq.~\bref{relation of C}.  Eq.~\bref{Cbar}
completes our derivation.  All the parameters are written in terms of
$\alpha_{n,1}^{-1}$ and ${C}^{(\epsilon)}_{n,1\mu}$ by eqs.~\bref{sol
for Cn}, \bref{sol alpha for cell}, \bref{summary alpha for pipe n} and
\bref{Cbar}.

Finally, we derive the relation between ${C}^{(\epsilon)}_{n,1\mu}$ and
$\bar{C}^{(\epsilon)}_{n,1\mu}$.  The relation allows us to take
$\bar{C}^{(\epsilon)}_{n,1\mu}$ and $\alpha^{-1}_{n,1}$ as independent
variables.
 
We find the relation
\begin{eqnarray}
C_{n,1\nu}^{(-\epsilon)}= \frac{\beta_c}{b_{\epsilon (-)^{n_1}1}(n)}
\Bigl(\alpha_{n,1}^{-1}- \alpha_{n,-\epsilon(-)^{n_{\nu}}\nu}^{\epsilon (-)^{n_1}1}
\Bigr)
\label{relation 1}
\end{eqnarray}
by making the replacements in eq.~\bref{def C 2} as $\lambda \rightarrow
1$, $\epsilon \rightarrow -\epsilon$ and $\sigma \rho \rightarrow
\epsilon (-)^{n_1}1$.  Rewriting \bref{sol for Cn} for $\mu=1$ with the
use of eq.~\bref{pipe, alpha and C bar}, we find that
\begin{eqnarray}
\frac{1}{\beta_p}\bar{C}^{(\epsilon)}_{n,1\nu} = \frac{1}{b_{\sigma \rho}(n)}
\alpha^{\sigma \rho}_{n,\epsilon(-)^{n_{1}}1}\Bigm|_{\sigma \rho =- \epsilon(-)^{n_{\nu}}\nu}
=-\frac{1}{b_{\sigma \rho}(n)}
\alpha_{n,- \epsilon(-)^{n_{\nu}}\nu}^{\sigma \rho}\Bigm|_{\sigma \rho =\epsilon(-)^{n_{1}}1}
\label{relation 2}
\end{eqnarray}
Removing $\alpha_{n,- \epsilon(-)^{n_{\nu}}\nu}^{\epsilon(-)^{n_{1}}1}$
from eqs.~\bref{relation 1} and \bref{relation 2}, we obtain
\begin{eqnarray}
C_{n,1\nu}^{(-\epsilon)}= \epsilon (-)^{n_1} \beta_c \alpha_{n,1}^{-1} + \frac{\beta_c}{\beta_p} \bar{C}_{n,1\nu}^{(\epsilon)}.
\label{C and Cbar}
\end{eqnarray}
The expressions of parameters in terms of
$\bar{C}^{(\epsilon)}_{n,1\mu}$ and $\alpha^{-1}_{n,1}$ at the end of
sect.~3 can be derived by using eq.~\bref{C and Cbar}.

%%%%%%%%%%%%%%%%%%%%%%%%%%%%%%%%%%%%%%%%%%%%%%%%%%%%%%%%%%%%%%%%%%%%%%
%%%%%%%%%%%%%%%%%%%%%%%%%%%%%%%%%%%%%%%%%%%%%%%%%%%%%%%%%%%%%%%%%%%%%%
\section{Reality of various coefficients}
%%%%%%%%%%%%%%%%%%%%%%%%%%%%%%%%%%%%%%%%%%%%%%%%%%%%%%%%%%%%%%%%%%%%%%
%%%%%%%%%%%%%%%%%%%%%%%%%%%%%%%%%%%%%%%%%%%%%%%%%%%%%%%%%%%%%%%%%%%%%%

Here we study the reality properties of various coefficients.  In
\cite{1st}, we noticed that the combination $(\alpha \cdot
\zeta)_{n,\sigma \mu}$ must be pure imaginary for the measure
invariance.  {}From the definition, the condition may be written as
\begin{eqnarray}
&{}&(\alpha \cdot \zeta)_{n,\sigma \mu}^{\dagger}
= U_{n,\sigma^{\prime} \rho}
\xi_{n+\sigma^{\prime} \rho}^{\dagger}U_{n,\sigma^{\prime} \rho}^{\dagger}
(\alpha_{n,\sigma \mu}^{\sigma^{\prime} \rho})^{*} 
\nonumber\\
&=&-(\alpha \cdot \zeta)_{n,\sigma \mu}
= -\alpha_{n,\sigma \mu}^{\sigma^{\prime} \rho} U_{n,\sigma^{\prime} \rho}
\xi_{n+\sigma^{\prime} \rho}U_{n,\sigma^{\prime} \rho}^{\dagger}.
\end{eqnarray}
Using eq.~\bref{xi reality}, we find the relation, 
\begin{eqnarray}
(\alpha_{n,\sigma \mu}^{\sigma^{\prime} \rho})^{*}
= (-)^{|n|+1}~\alpha_{n,\sigma \mu}^{\sigma^{\prime} \rho}.
\label{phase a}
\end{eqnarray}

The consistency of eqs. \bref{fermion tf} and \bref{xi reality} implies that
\begin{eqnarray}
(C_{n,\mu\nu}^{\delta,\epsilon})^* =  (-)^{|n|} C_{n,\mu\nu}^{\delta,\epsilon}.
\label{reality of C}
\end{eqnarray}
As for the link variable, it is easy to confirm that the notation
$U_{n+{\hat \mu}, -\mu} \equiv U_{n,\mu}^{\dagger}$ is consistent with
eq.~\bref{One cell del U}, when $(\alpha \cdot \zeta)_{n,\sigma \mu}$
is pure imaginary.  

%%%%%%%%%%%%%%%%%%%%%%%%%%%%%%%%%%%%%%%%%%%%%%%%%%%%%%%%%%%%%%%%%%%%%%
%%%%%%%%%%%%%%%%%%%%%%%%%%%%%%%%%%%%%%%%%%%%%%%%%%%%%%%%%%%%%%%%%%%%%%
\section{The Majorana staggered fermion}
%%%%%%%%%%%%%%%%%%%%%%%%%%%%%%%%%%%%%%%%%%%%%%%%%%%%%%%%%%%%%%%%%%%%%%
%%%%%%%%%%%%%%%%%%%%%%%%%%%%%%%%%%%%%%%%%%%%%%%%%%%%%%%%%%%%%%%%%%%%%%

In the first two subsections we describe how our fermion is related to
Majorana fermions naively discretized on lattice.  In the last
subsection, we discuss the reconstruction problem, ie, how we could
obtain the Majorana fermion at the naive continuum limit.

%%%%%%%%%%%%%%%%%%%%%%%%%%%%%%%%%%%%%%%%%%%%%%%%%%%%%%%%%%%%%%%%%%%%%
\subsection{The Majorana staggered fermion}
%%%%%%%%%%%%%%%%%%%%%%%%%%%%%%%%%%%%%%%%%%%%%%%%%%%%%%%%%%%%%%%%%%%%%%

First take a free Dirac fermion action naively discretized on the
lattice,
\begin{eqnarray}
  \sum_{n,\mu}\bar{\psi}_{n}\gamma_{\mu}
  \left(\frac{\psi_{n+\hmu}-\psi_{n-\hmu}}{2}\right).
\label{naive dis Dirac}
\end{eqnarray}
As for the $\gamma_{\mu}$, we follow Kugo's
notation\cite{Kugo}\cite{KugoTownsend}.  In particular, $\gamma_{\mu}$
are hermitian and thus unitary in the Euclidean space.

We impose the Majorana condition on $\psi$ as
\begin{eqnarray}
\psi_n = \psi_n^c \equiv C {\bar \psi}_n^T.
\label{Majorana condition}
\end{eqnarray}
The charge conjugation matrix $C$ has the properties
\begin{eqnarray}
C^{-1} \gamma_{\mu} C = \eta^{\prime} \gamma_{\mu}^T,~~~~~
C^T = \varepsilon^{\prime} C.
\label{charge conjugation matrix}
\end{eqnarray}
with sign factors $\eta^{\prime}$ and $\varepsilon^{\prime}$.  In
Ref.~\cite{KugoTownsend} one finds the table relating the space
dimension to the allowed values for these factors.  When the fermion
carries the flavor indices, we may impose other Majorana conditions.
That possibility is not discussed in this paper.

Following the standard procedure\cite{Rothe}, we rewrite the action in
terms of the new variables $\omega_n$ defined as
\begin{eqnarray}
 \psi_{n} \equiv V_{n}\omega_{n} \ , \ \ 
  \omega_{n}=\left(\begin{array}{c}\xi_{n}^{(1)}\\
  \xi_{n}^{(2)} \\ \vdots \end{array}\right)   
\end{eqnarray}
where
\begin{eqnarray}
 V_{n} \equiv \gamma_{1}^{n_{1}}\gamma_{2}^{n_{2}} \cdots \gamma_{d}^{n_{d}}.
\end{eqnarray}
The fermion action is now expressed as 
\begin{eqnarray}
\sum_{n,\mu}\bar{\psi}_{n}\gamma_{\mu}
\left(\frac{\psi_{n+\hmu}-\psi_{n-\hmu}}{2}\right)
&=& \frac{\varepsilon^{\prime}}{2} \sum_{n,\mu>0}
(\eta^{\prime})^{|n|}\eta_{\mu}(n) 
\Bigl[ \omega_n^T C^{-1} (\omega_{n+{\hat \mu}}- 
\omega_{n-{\hat \mu}}) \Bigr]
\nonumber\\%
&=& \frac{1}{2}(\varepsilon^{\prime}+\eta^{\prime}) 
\sum_{n,\mu>0}(\eta^{\prime})^{|n|}\eta_{\mu}(n)~\omega_n^T C^{-1}
\omega_{n+{\hat \mu}}.
\label{omega action}
\end{eqnarray}
In the second equality, the use has been made of the relation
 \begin{eqnarray*}
  {\omega_{n}}^{T}C^{-1}\omega_{n+\hmu}= - {\omega_{n+\hmu}}^{T}(C^{-1})^{T}\omega_{n}
  &\longrightarrow& - {\omega_{n}}^{T}(C^{-1})^{T}\omega_{n-\hmu}.\\
  &\scriptstyle{n+\hmu \rightarrow n}& 
 \end{eqnarray*}
{}From eq. \bref{omega action}, it is easy to see that the lattice
Majorana fermion is allowed if
\begin{eqnarray}
  \varepsilon^{\prime}\eta^{\prime}= +1.
\label{the condition}
\end{eqnarray}

Following the standard procedure for the staggered Dirac fermion, we
would like to reduce the number of degrees of freedom on a site.
Suppose we have only one independent component in $\omega_n$, ie,
\begin{eqnarray}
  \omega_{n}=\xi_{n} u_{0} 
\label{one component}
\end{eqnarray}
with $u_0$ as a constant spinor.  Then we have
\begin{eqnarray}
  {\omega_{n}}^{T}C^{-1}\omega_{n+\hmu}
  =\xi_{n}\xi_{n+\hmu}\Bigl({u_{0}}^{T}C^{-1}u_{0}\Bigr)
  ~~\propto~~ \left\{\begin{array}{cc} = 0 & ~~~(\varepsilon^{\prime}=-1) \\
  \neq 0 & ~~~(\varepsilon^{\prime}=+1) \end{array}\right.
\end{eqnarray}
owing to the property of the charge conjugation matrix.  Therefore we
may conclude that the expression \bref{one component} is allowed only for the
case of $\varepsilon^{\prime}=+1$ (and $\eta^{\prime}=+1$).  In this
case, the action \bref{omega action} becomes the fermion action
\bref{fermion action}, up to an irrelevant normalization coefficient,
\begin{eqnarray}
S_{\rm naive} = \Bigl({u_{0}}^{T}C^{-1}u_{0}\Bigr) \sum_{n,\mu} \eta_{\mu}(n) \xi_n
\xi_{n+{\hat \mu}}.
\label{naive action}
\end{eqnarray}

We can derive the reality condition on $\xi_n$,
\begin{eqnarray}
\xi_n^{\dagger} = \frac{(u_0^T C^{-1} u_0)}{(u_0^{\dagger}\gamma_{D+1}u_0)} (-\eta^{\prime})^{|n|} \xi_n,
\label{reality on xi}
\end{eqnarray}
from the Majorana condition \bref{Majorana condition} and the definition
of ${\bar \psi}$
\begin{eqnarray}
{\bar \psi} \equiv \psi^{\dagger} \gamma_{D+1}.
\label{def bar psi}
\end{eqnarray}
Though $\eta^{\prime}=1$, we leave eq.~\bref{reality on xi} as it is to
remember where the sign factor comes from.

The matrix $\gamma_{D+1}$, given in eq.~\bref{gamma D+1 1}, is available
only for even dimensional space.  So the definition \bref{def bar psi} of ${\bar \psi}$ is
valid for that case.  The odd dimensional cases require separate
consideration, that will not be discussed in this paper.

{}From the way $\xi_n$ and $u_0$ appeared in eq.~\bref{one component}, clearly
we have the freedom to multiply a phase factor and its inverse to $\xi_n$
and $u_0$ simultaneously, In particular, we may make positive the
coefficient of the action \bref{naive action} or the numerator of
eq.~\bref{reality on xi}.  Therefore the fermion action and the reality
condition on the fermionic variable can be chosen as
\begin{eqnarray}
S_{\rm naive} = \sum_{n,\mu} \eta_{\mu}(n) \xi_n
\xi_{n+{\hat \mu}},
\label{naive action 2}
\end{eqnarray}
and 
\begin{eqnarray}
\xi_n^{\dagger} = (-\eta^{\prime})^{|n|} \xi_n.
\label{reality on xi 2}
\end{eqnarray}
The denominator of eq.~\bref{reality on xi},
${(u_0^{\dagger}\gamma_{D+1}u_0)}$, is obviously real.  Therefore the
coefficient on the rhs of eq.~\bref{reality on xi} is either $+1$ or
$-1$.  In writing eqs.~\bref{naive action 2} and \bref{reality on xi 2}, 
we ignored this possible sign factor.

This is how we reached the fermion action and the reality condition
used in this paper.  

In the next subsection, we consider the reconstruction problem starting
from the action \bref{naive action 2}.  Soon we realize that the
reduction from eq.~\bref{naive dis Dirac} and the reconstruction from
eq.~\bref{naive action 2} give rise different conditions on the sign
factors $\varepsilon^{\prime}$ and $\eta^{\prime}$.  This is no
surprise, since we reduced the number of degrees of freedom as \bref{one
component} in the middle of the reduction process.  In considering the
proper continuum limit, the reconstruction problem and its results are
far more important than the reduction.

%%%%%%%%%%%%%%%%%%%%%%%%%%%%%%%%%%%%%%%%%%%%%%%%%%%%%%%%%%%%%%%%%%%%%
%%%%%%%%%%%%%%%%%%%%%%%%%%%%%%%%%%%%%%%%%%%%%%%%%%%%%%%%%%%%%%%%%%%%%
%%%%%%%%%%%%%%%%%%%%%%%%%%%%%%%%%%%%%%%%%%%%%%%%%%%%%%%%%%%%%%%%%%%%%
%%%%%%%%%%%%%%%%%%%%%%%%%%%%%%%%%%%%%%%%%%%%%%%%%%%%%%%%%%%%%%%%%%%%%
\subsection{Reconstruction problem for the free Majorana fermion}
%%%%%%%%%%%%%%%%%%%%%%%%%%%%%%%%%%%%%%%%%%%%%%%%%%%%%%%%%%%%%%%%%%%%%

In this subsection we reconstruct the (free) Majorana fermion on the
lattice with the spacing $2a$.\footnote{We follow the discussion for the
staggered Dirac fermion given in Ref. \cite{Rothe}.  The important
difference is in the presence of the Majorana condition.}  We start from
the action \bref{fermion action}
\begin{eqnarray}
S_f= \sum_{n,\mu} \eta_{\mu}(n) {\rm tr}(\xi_n \xi_{n+{\hat \mu}})=\frac{1}{2} \sum_{n,\mu>0} \eta_{\mu}(n)
{\rm tr}(\xi_n \xi_{n+{\hat \mu}}-\xi_n \xi_{n-{\hat \mu}}).
\label{Sf action}
\end{eqnarray}
The action \bref{Sf action} is real with the reality condition assigned
on the fermion $\xi_n^{\dagger} = (-)^{|n|}\xi_n$ (see appendix B).
Note that the coefficient in the fermion action is taken as
$b_{\mu}(n)=\eta_{\mu}(n)$.

We understand that the degrees of freedom are distributed in a
hypercube attached to the reference point $2 N=2 \times (N_1, N_2,
\cdots N_D)$, where $N_{\mu}$ are the integers.  So let us write the
site coordinates as 
\begin{eqnarray} 
n_{\mu} \equiv 2 N_{\mu}+ r_{\mu}.
\end{eqnarray}
Here $r_{\mu}$, being the relative coordinate, takes the values of $0$
or $1$.  Since $\eta_{\mu}(n)=\eta_{\mu}(r)$, the fermion action may
be rewritten as 
\begin{eqnarray}
S_f = \frac{1}{2}\sum_{N,r}\sum_{\mu>0} \eta_{\mu}(r) 
{\rm tr}\Bigl(\xi_{2N+r} \Bigl[\xi_{{2N+r+{\hat \mu}}}-\xi_{2N+r-{\hat \mu}}\Bigr]\Bigr).
\label{Sf action 2}
\end{eqnarray}
Collecting the $2^D$-fields in the hypercube with the reference coordinate
$2N$, we define
\begin{eqnarray}
\xi_{r}(N) \equiv \xi_{2N+r}.
\end{eqnarray}
It is easy to find the following relations
\begin{eqnarray}
\xi_{{2N+r+{\hat \mu}}} &=& \sum_{r^{\prime}}\Bigl[
\delta_{r+{\hat \mu},r^{\prime}} \xi_{r^{\prime}}(N)
+\delta_{r-{\hat \mu},r^{\prime}} \xi_{r^{\prime}}(N+{\hat \mu})  \Bigr],
\nonumber\\
\xi_{{2N+r-{\hat \mu}}} &=& \sum_{r^{\prime}}\Bigl[
\delta_{r-{\hat \mu},r^{\prime}} \xi_{r^{\prime}}(N)
+\delta_{r+{\hat \mu},r^{\prime}} \xi_{r^{\prime}}(N-{\hat \mu})  \Bigr].
\label{xi relation}
\end{eqnarray}
So we have
\begin{eqnarray}
\xi_{{2N+r+{\hat \mu}}}-\xi_{2N+r-{\hat \mu}}
\equiv \sum_{r^{\prime}}\Bigl[
\delta_{r+{\hat \mu},r^{\prime}} {\hat \partial}_{\mu}^L \xi_{r^{\prime}}(N)
+\delta_{r-{\hat \mu},r^{\prime}} {\hat \partial}_{\mu}^R \xi_{r^{\prime}}(N)
\Bigr]
\label{diff}
\end{eqnarray}
where we have introduced derivatives with respect to the lattice spacing $2a$,
\begin{eqnarray}
{\hat \partial}_{\mu}^L \xi_{r^{\prime}}(N) \equiv 
\xi_{r^{\prime}}(N) - \xi_{r^{\prime}}(N-{\hat \mu}),
~~~~~
{\hat \partial}_{\mu}^R \xi_{r^{\prime}}(N) \equiv
\xi_{r^{\prime}}(N+{\hat \mu}) - \xi_{r^{\prime}}(N). 
\label{derivative1}
\end{eqnarray}
Substituting eq.~\bref{diff} into eq.~\bref{fermion action}, we obtain
\begin{eqnarray}
S_f = \frac{1}{2} \sum_{N, r, r^{\prime}} \sum_{\mu >0} 
\Bigl[
\xi_{r}(N) \Bigl(
\Gamma_{r r^{\prime}}^{\mu} {\hat \partial}_{\mu}
+\frac{1}{2}
\Gamma_{r r^{\prime}}^{5 \mu}
{\hat \square}_{\mu}
\Bigr) \xi_{r^{\prime}}(N) \Bigr]
\label{xi 2a action}
\end{eqnarray}
where
\begin{eqnarray}
&{}&\Gamma_{r r^{\prime}}^{\mu} \equiv
(\delta_{r+{\hat \mu},r^{\prime}}+\delta_{r-{\hat \mu},r^{\prime}})\eta_{\mu}(r),
~~~
\Gamma_{r r^{\prime}}^{5 \mu} \equiv
(\delta_{r-{\hat \mu},r^{\prime}}-\delta_{r+{\hat \mu},r^{\prime}})\eta_{\mu}(r),
\nonumber\\
&{}&~~~~~~~{\hat \partial}_{\mu}  \xi_{r^{\prime}}(N) \equiv
\frac{1}{2} \Bigl( 
\xi_{r^{\prime}}(N+{\hat \mu}) -
\xi_{r^{\prime}}(N-{\hat \mu}) \Bigr), 
\nonumber\\
&{}&~~~~~~~{\hat \square}_{\mu} \xi_{r^{\prime}}(N) \equiv \xi_{r^{\prime}}(N+{\hat \mu}) +
\xi_{r^{\prime}}(N-{\hat \mu}) - 2 \xi_{r^{\prime}}(N).
\label{definition1}
\end{eqnarray}
Obviously only the first term survives in the naive continuum
limit.  Therefore we consider the first term given in
eq. \bref{lattice Sf2}
\begin{eqnarray}
S_f = \frac{1}{2} \sum_{N,\mu>0,r}\eta_{\mu}(r) \xi_r(N){\hat \partial}_{\mu}
\Bigl[ \xi_{r+\hmu}(N) + \xi_{r-\hmu}(N) \Bigr] + \cdots.
\label{lattice Sf2a}
\end{eqnarray}

We would like to further rewrite the above action in terms of the
reconstructed fermion
\begin{eqnarray}
{\hat \psi}_{\alpha i}(N) &\equiv& {\cal N}_0 \sum_{r} (V_r)_{\alpha i} \xi_r(N),
\label{hat psi}
\end{eqnarray}
The coefficient ${\cal N}_0$ is a normalization constant to be fixed
appropriately.  Note that $V_r$ has the properties
\begin{eqnarray}
\sum_{r}(V_r^{\dagger})_{i\alpha}(V_r)_{\beta j}&=2^{\frac{D}{2}}&\delta_{\alpha\beta} \delta_{ij},~~~~~
V_r^{\dagger}=d_rV_r \nonumber\\
{\rm tr}(V_r^{\dagger} V_{r^{\prime}}) &=& \delta_{r r^{\prime}} 
2^{\frac{D}{2}},~~~~~
V_{r\pm{\hat \mu}}=\eta_{\mu}(r) \gamma_{\mu} V_r,
\label{properties of Vr}
\end{eqnarray}
where $d_r=(-)^{|r|(|r|-1)/2}$ is a sign factor.
Reversing the expression \bref{hat psi}, we have
\begin{eqnarray}
\xi_r(N) = \frac{1}{\cal N}_0 {\rm tr} \Bigl[V_r^{\dagger} {\hat \psi}(N)\Bigr].
\label{xi expression}
\end{eqnarray} 

We would like to write the action \bref{xi 2a action} in terms of ${\hat
\psi}$.  For that purpose, it is useful to remember how it goes for the
Dirac fermion.  In that case we have ${\bar \xi}_r$ and ${\xi}_r$ which
are related to the reconstructed fermions as \bref{xi and psi}.  Using the
relation,
\begin{eqnarray}
\sum_{r}\eta_{\mu}(r) {\bar \xi}_r(N){\hat \partial}_{\mu} \xi_{r\pm\hmu}(N)
&=&({\cal N}_0)^{-2} \sum_r \eta_{\mu}(r)
{\rm tr}\Bigl[{\bar {\hat \psi}}(N) V_r\Bigr]~
{\hat {\partial}}_{\mu}
{\rm tr}\Bigl[V_{r\pm\hmu}^{\dagger} {\hat \psi}(N)\Bigr]\nonumber\\
&=&({\cal N}_0)^{-2} 2^{D/2} 
{\bar {\hat \psi}}(N)_{i \alpha} 
(\gamma_{\mu})_{\alpha \beta}{\hat \partial_{\mu}}
{\hat \psi}(N)_{\beta i},
\label{Dirac Staggered fermion}
\end{eqnarray}
the action $S_{\rm Dirac}$ of the (Dirac) staggered fermion can be obtained
as
\begin{eqnarray}
S_{\rm Dirac} = \sum_{N,\mu} {\rm tr} \Bigl[{\bar {\hat \psi}}(N) 
(\gamma_{\mu}){\hat \partial_{\mu}}
{\hat \psi}(N)\Bigr]+ \cdots.
\label{Dirac action}
\end{eqnarray}
Here the normalization constant is chosen as ${\cal N}_0=2^{D/4}$.
 
If we expect to have the relation similar to eq.~\bref{Dirac Staggered
fermion} for the Majorana staggered fermion, $\xi_r(N)$ must be related
to both ${\hat \psi}(N)$ and ${\bar {\hat \psi}}(N)$ as
\begin{eqnarray}
\xi_r(N) = \frac{1}{{\cal N}_0} {\rm tr} \Bigl[
V_r^{\dagger} {\hat \psi}(N)\Bigr]
= \frac{1}{{\cal N}_0} {\rm tr} \Bigl[
{\bar {\hat \psi}}(N) V_r\Bigr].
\label{Maj xi and psi}
\end{eqnarray}
For the second equality to hold, we find the Majorana condition on ${\hat 
\psi}$ with $\eta^{\prime}=1$
\begin{eqnarray} 
{\hat \psi} = C {\bar {\hat \psi}}^TC^{-1}.
\label{Usp Majorana}
\end{eqnarray} 
The restriction $\eta^{\prime}=1$ may be easily read off from the following
relation,
\begin{eqnarray} 
{\rm tr}\Bigl[{\hat \psi}V_r^{\dagger}\Bigr]
= {\rm tr}\Bigl[{\bar {\hat \psi}}^T C^{-1} \gamma_D^{r_D} \cdots \gamma_1^{r_1}C\Bigr]
= {\rm tr}\Bigl[{\bar {\hat \psi}}^T (\eta^{\prime})^{|r|} V_r^T\Bigr]
= (\eta^{\prime})^{|r|}{\rm tr} \Bigl[{\bar {\hat \psi}} V_r \Bigr].
\nonumber
\end{eqnarray} 

Eq.~\bref{Maj xi and psi} and the reality condition on $\xi_n$, ie, $\xi_n^{\dagger} =
(-)^{|n|}\xi_n$, imply the relation
\begin{eqnarray}
{\bar {\hat \psi}}= \gamma_{D+1}^{\dagger} {\hat \psi}^{\dagger} \gamma_{D+1},
\label{psi bar and psi dagger}
\end{eqnarray}
where $\gamma_{D+1}$ is a hermitian matrix defined in eq.~\bref{gamma
D+1 1}.  Remember that, for a Euclidean theory, ${\bar \psi} \equiv
\psi^{\dagger} \gamma_{D+1}$ is the appropriate relation to keep the
reality of the action ${\bar \psi}(\gamma_{\mu}\partial_{\mu}+m)\psi$.
Eq.~\bref{psi bar and psi dagger} is a multi-flavored extension of this
relation.
The matrix $\gamma_{D+1}$ is defined only for even dimensional spaces
and we have to consider the odd dimension separately.  That is, however,
beyond the scope of the present paper.

In Ref.\cite{KugoTownsend}, we find the list of Majorana fermions
realized by the condition given in eq.~\bref{Usp Majorana}, as already
presented in the sect.~3.

%%%%%%%%%%%%%%%%%%%%%%%%%%%%%%%%%%%%%%%%%%%%%%%%%%%%%%%%%%%%%%%%%%%%%%
%%%%%%%%%%%%%%%%%%%%%%%%%%%%%%%%%%%%%%%%%%%%%%%%%%%%%%%%%%%%%%%%%%%%%%
\section{Uniqueness of the models}
%%%%%%%%%%%%%%%%%%%%%%%%%%%%%%%%%%%%%%%%%%%%%%%%%%%%%%%%%%%%%%%%%%%%%%
%%%%%%%%%%%%%%%%%%%%%%%%%%%%%%%%%%%%%%%%%%%%%%%%%%%%%%%%%%%%%%%%%%%%%%

%%%%%%%%%%%%%%%%%%%%%%%%%%%%%%%%%%%%%%%%%%%%%%%%%%%%%%%%%%%%%%%%%%%%%%
%%%%%%%%%%%%%%%%%%%%%%%%%%%%%%%%%%%%%%%%%%%%%%%%%%%%%%%%%%%%%%%%%%%%%%
\begin{floatingfigure}
%%%%%%%%%%%%%%%%%%%%%%%%%%%%%%%%%%%%%%%%
%%%  Figure (3-Dim. Planes)          %%%
%%%%%%%%%%%%%%%%%%%%%%%%%%%%%%%%%%%%%%%%
%% ---------------------------------------------------------------- %%
{\setlength{\unitlength}{0.2mm}
\begin{picture}(400,320)(0,0)
 \thicklines 
 \put(50,100){\vector(1,0){270}}    
 \put(152,64){\vector(-4,-3){32}} 
% \put(280,100){\vector(1,0){60}}  
 \put(200,100){\vector(0,1){200}} 
 \put(152,64){\line(0,1){80}}
 \put(232,64){\line(0,1){80}}
 \put(280,100){\line(0,1){80}}
 \put(200,100){\path(0,0)(80,0)(32,-36)(-48,-36)(0,0)}
 \put(200,180){\path(0,0)(80,0)(32,-36)(-48,-36)(0,0)}
 \put(200,260){\path(0,0)(80,0)(32,-36)(-48,-36)(0,0)}
 \put(62,64){\path(0,0)(0,80)(48,116)(48,36)(0,0)}
 \put(216,122){\circle*{8}}
% \put(224,114){\line(1,-1){60}}    
 \put(284,54){\vector(-1,1){60}}    
% \put(205,115){\huge $Q$}
 \put(290,40){\large $Q$}
 \put(200,75){\Large $F_{0}$}  \put(245,115){\Large $F_{0}^{\prime}$}
 \put(205,155){\Large $F_{1}$} \put(165,115){\Large $F_{1}^{\prime}$}
 \put(205,235){\Large $F_{2}$} \put(75,115){\Large $F_{2}^{\prime}$}
 \put(90,30){\Large $x_{1}$}
 \put(330,90){\Large $x_{2}$}
 \put(205,305){\Large $x_{3}$}
\end{picture}}
%%%%%%%%%%%%%%%%%%%%%%%%%%%%%%%%%%%%%%%%
 \caption{This figure shows how the face $F_0$ is related to other faces
by four conditions listed in the text.}
\end{floatingfigure}
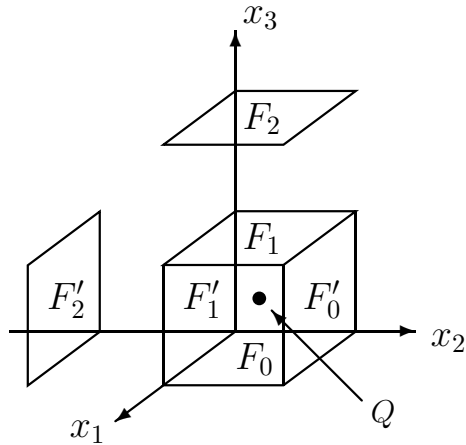
%%%%%%%%%%%%%%%%%%%%%%%%%%%%%%%%%%%%%%%%

Here we find out how we can classify the plaquettes in the entire space
so that the following four conditions are satisfied: 1) Ichimatsu
pattern on any two dimensional surface\footnote{We do not necessarily
assume that the plaquette patterns on two parallel surfaces are the
same.}  ; 2) invariance under mod 2 translations; 3) symmetry under
${\pi}/2$ rotations; 4) the reflection positivity.
The cell, the pipe and the mixed models are uniquely selected out by the 
first three conditions; the models have the reflection positivity to
allow an appropriate continuum limit, as already discussed in sect.~5.2.

{}From the condition 1), we are to classify plaquettes into shaded and
unshaded sets on every two dimensional surface.  In the following, we
show that the assignment is uniquely determined by two other
conditions.  Based on this pattern, we construct the cell, pipe and
mixed models.  Collecting shaded (unshaded) plaquettes, we obtain the
cell (pipe) model.  For the mixed model, we give the coefficients
$\beta_c$ ($\beta_p$) to shaded (unshaded) plaquettes.  In this sense,
the cell, pipe and mixed models are uniquely selected out by the conditions.

The three dimensional case suffices to illustrate our consideration.
Take a two dimensional plane and choose the coordinate system so that
the plane is the $x_1 x_2$-plane with $x_D \equiv x_3=0$.  On this
plane, we take the plaquette closest to the origin in the first quadrant
and make it shaded: the plaquette is $F_0$ in Fig.~4.  Remember that
this choice completely determines the Ichimatsu pattern on the plane.

Next we find out how we may realize rotations consistently with the
Ichimatsu pattern.  Take again the $x_1 x_2$-plane in the last paragraph
and consider the $\pi /2$ rotation of this plane which does not affect
the Ichimatsu pattern on the plane.  Clearly the center of the rotation
must be located at the center of a plaquette.  Its rotational axis
passes through the center of a hypercube, eg, $Q \equiv (1/2,~1/2,~1/2)$
shown in Fig.~4.  Consider two other axes through the same point $Q$ and
parallel to $x_1$ and $x_2$.  We are to arrange the pattern in the
entire space so that the $\pi /2$ rotations around these axes do not
change the pattern.

Although the reflection is not necessary for the present discussion, it
is appropriate to mention how the reflection plane should be chosen.
The reflection must be consistent with the Ichimatsu patterns in the
entire space.  There are two types of planes, perpendicular to the
$x_3$-axis, which respect the lattice structure.  One is the
$x_1x_2$-plane and the other is the plane that contains the point $Q$.
Obviously the reflection with respect to the former {\it does change} the
Ichimatsu pattern on the $x_1x_3$-plane.  Therefore the reflection plane
must contain the center of a hypercube.

In Fig.~4, the face $F_0$ is a shaded plaquette.  Using twice the $\pi
/2$ rotation around the $x_1$ axis, we find that $F_1$ must also be
shaded.  From the invariance under the translation by mod 2, $F_2$ is
also shaded.  These results completely determine the Ichimatsu patterns
on planes parallel to the $x_1 x_2$-plane.  By the $\pi /2$ rotation
around the $x_1$ axis, the plaquettes $F_0$, $F_1$ and $F_2$ are brought
to those with primes, and the Ichimatsu patterns parallel to the $x_1
x_3$-plane are determined.  Obviously, rotations around another axis
give the unique and complete classification of all the plaquettes into
the shaded and unshaded set.  This proves our claim for the three
dimensional case.  It must be obvious that the proof may be easily
extended to any dimension.

%%%%%%%%%%%%%%%%%%%%%%%%%%%%%%%%%%%%%%%%%%%%%%%%%%%%%%%%%%%%%%%%%%%%%%
%%%%%%%%%%%%%%%%%%%%%%%%%%%%%%%%%%%%%%%%%%%%%%%%%%%%%%%%%%%%%%%%%%%%%%

%%%%%%%%%%%%%%%%%%%%%%%%%%%%%%%%%%%%%%%%%%%%%%%%%%%%
%%%%%%%%%%%%%%%%%%%%%%%%%%%%%%%%%%%%%%%%%%%%%%%%%%%%
%%%%%%%%%%%%%%%%%%%%%%%%%%%%%%%%%%%%%%%%%%%%%%%%%%%%

%%%%%%%%%%%%%%%%%%%%%%%%%%%%%%%%%%%%%%%%%%%%%%%%%%%%
\end{document}